# Voltage-Based Unsupervised Learning Framework for Bridge Damage Detection in Simultaneous Energy Harvesting and Sensing Systems

S. Yao[a], P. Peralta-Braz[a], A. Calderon Hurtado[a], R. Das[b], M. M. Alamdari[1a], E. Atroshchenko[a]

[a]*School of Civil and Environmental Engineering, University of New South Wales, Sydney, Australia*
[b]*School of Engineering, RMIT University, Melbourne, Australia*

**Abstract**

In this study, piezoelectric energy harvesters (PEHs) are designed to offer a dual functionality in Structural Health Monitoring (SHM): harvesting electric power from bridge vibrations while serving as intrinsic damage sensors. This strategy utilises the voltage signal directly as the sensing input, eliminating the need for traditional sensing modules, thereby achieving a reduction in system complexity and energy consumption. This study also presents a bi-objective optimisation framework that maximises both power output and damage-detection accuracy of a PEH modelled as a composite cantilevered Kirchhoff–Love plate. Voltage responses under realistic bridge inputs are predicted via isogeometric analysis. We validate our approach in two scenarios: a numerical vehicle–bridge interaction model and a laboratory-scale beam test using a toy car, each evaluated under both healthy and damaged states. Unsupervised damage detection is achieved with a convolutional variational autoencoder (CVAE) trained solely on healthy voltage signatures. We then apply the NSGA-II algorithm to explore trade-offs between energy yield and sensing precision, conducting parametric studies on damage severity, location, and harvester geometry. Results reveal that optimised PEHs not only act as an effective filter and sensing component, but also outperform traditional acceleration-based sensing by enhancing damage detection accuracy by 13% while reducing energy consumption by 98%. The case with multi-parameter design space highlights the importance of bi-objective optimisation, due to observed deviations in performance of PEHs even under resonant conditions. These findings demonstrate the feasibility of replacing traditional sensors with lightweight, self-powered PEHs and pave the way for sustainable, Simultaneous Energy Harvesting and Sensing (SEHS) systems.

*Keywords:* Piezoelectric energy harvester, unsupervised learning, isogeometric analysis, simultaneous energy harvesting and sensing, vehicle-bridge interaction, multi-objective optimisation

[1]Corresponding author, m.makkialamdari@unsw.edu.au

# 1. Introduction

Infrastructure, such as bridges, plays a critical role in societal and economic development [1]. However, these structures are susceptible to unpredictable damage, such as the formation of small cracks, which, if untreated, can compromise operational efficiency and pose significant safety risks [2]. With the development of the Internet of Things (IoT), various wireless sensors, such as accelerometers and strain gauges, are widely deployed in infrastructure to perform structural health monitoring (SHM) and facilitate management and maintenance operations [3]. SHM systems utilise such technologies to provide real-time data collection for active monitoring, aiding in reporting and characterising the structural behaviour and performance of infrastructure [4]. However, a significant limitation of IoT sensing architecture is the reliance on a stable power supply. In existing SHM systems, the lack of continuous power and excessive energy dissipation often lead to data loss or even monitoring interruptions, posing serious challenges to the reliability and sustainability of long-term sensing[5]. Addressing this problem requires a dual approach: maximising energy availability and minimising power consumption.

In terms of energy supply, compared to wired connections or battery packs that require frequent maintenance and replacement, renewable sources such as solar and wind offer continuous power and environmental benefits [6]. However, the inaccessibility of traditional renewable technologies, such as the uncontrollability of photovoltaic cells in bad weather, defeats the purpose of uninterrupted sensing [7]. Given the ubiquity of ambient mechanical vibrations in civil infrastructure, piezoelectric energy harvesters (PEHs) present applicable solutions due to their ability to efficiently convert vibrations into electrical energy [8]. Nevertheless, the inherently low power output of PEHs remains a major constraint. Therefore, extensive research has focused on optimising PEH performance through metamaterials, non-linear mechanisms, and structural designs ([9], [10], [11], [12]). For example, Jamshiddoust et al. designed an auxetic composite plate structure incorporating both $d_{31}$ and $d_{32}$ piezoelectric coupling effects, achieving strain amplification and improving performance by 30% [13]. To further enhance the energy harvesting capability and achieve more sustainable power supply in the context of real bridge SHM systems, Yao et al. employed an optimisation framework based on structural design and particle swarm optimisation (PSO) to identify the best PEH locations on a real bridge, resulting in up to 3.6 times higher energy output depending on the installation position and traffic intensity [14]. In another approach, Huang et al. proposed a magnetoelastic bi-stable vibration absorber with tunable stiffness, which enabled broadband energy harvesting under weak ambient excitations through enhanced non-linear dynamic response [15]. These studies demonstrate the effectiveness of structural innovations and optimisation strategies in improving PEHs' energy output, making them increasingly capable of powering sensor nodes for SHM. For instance, Wang et al. developed a self-powered temperature monitoring device by integrating PEH and temperature sensing modules, which is sufficient to perform 140 measurements per hour [16]. While this highlights the potential of PEH-based autonomous sensing systems, the implementation involves a complex architecture with energy storage components, regulation circuits, and additional sensing modules. Such configurations increase the overall power demand and limit both the long-term operation and deployment density of the sensing system [17]. This motivates the exploration of strategies to minimise energy loss under a constrained power supply. In this context, replacing sens-ing modules with PEHs themselves emerges as a promising approach to significantly reduce system complexity and energy consumption.

Recently, the growing interest in PEHs as multifunctional components is attributed to their additional role as sensors, where the generated electrical signals can be directly exploited to extract environmental information. For instance, wearable piezoelectric devices have shown potential in detecting behavioural characteristics in human activities, demonstrating capabilities comparable to accelerometers in distinguishing different activity patterns such as walking, running, and cycling [18]. The integration of PEHs as sensors highlights their dual functionality, enabling both activity monitoring and physiological sensing. By utilising the high sensitivity of the piezoelectric effect, PEHs as medical sensors are able to effectively reflect physiological characteristics through voltage signals, like capturing subtle variations in organs and blood pressure in an implantable form [19]. Beyond medical applications, the deployment of PEHs as sensors in engineered environments including industrial and civil facilities has also shown superior applicability, where electrical output serves as a means of conveying sensing information. For example, the characteristics of ambient wind and water flow can be revealed through the peak electrical output of piezoelectric film devices, offering an alternative approach for engineering inspections, such as pipeline monitoring [20]. Additionally, the use of PEHs installed on vehicles as active sensing elements for traffic information feedback is feasible, as the voltage distribution during energy harvesting reflects the dynamic passage characteristics of vehicles such as speed [21]. Regarding SHM of infrastructure, there is evidence that the difference in output energy of PEHs in bridges corresponds to different damage states [22]. Moreover, signal processing and analysis of voltage output data enable effective SHM. For instance, Paul et al. demonstrated the potential of PEHs in identifying bridge scour damage by conducting modal analysis on voltage signals generated under a vibrating bridge [23].

Accordingly, the prospect of utilising sustainable energy harvesting devices to provide self-powered sensing services for the IoT is achievable through the implementation of PEHs as Simultaneous Energy Harvesting and Sensing (SEHS) modules. First, the dual functionality of PEHs dispenses with the use of active sensors, thereby simplifying the complexity and reducing the cost of sensing systems. For example, in human activity monitoring applications, a single insole-shaped PEH is capable of performing both energy harvesting and gait recognition, eliminating the need for hardware dedicated to individual sensing tasks, such as accelerometers [24]. Moreover, integrating energy harvesting and sensing reduces power consumption associated with traditional sensing components, thus promoting the long-term and sustainable operation of the entire system [25]. In the use of cantilever PEHs as smart home sensors, Rueda et al. found that longer devices are beneficial for the accurate detection of floor deformations. In contrast, the optimisation criteria for power output may be different, as PEH designs that achieve higher energy output often exhibit lower sensitivity in sensing applications [26]. Thus, the functional trade-offs of using PEHs as both sensors and harvesters raise relevant design and optimisation challenges. In order to explore the efficient operation of PEHs in SEHS systems, Peralta et al. performed a rigorous bi-objective design optimisation of a PEH with the aim to maximise both, energy harvesting capability and sensing accuracy in the context of vehicle speed classification [27]. Based on the SHM data from a real cable stayed bridge, they showed that the optimal solutions form a Pareto front.

However, despite the progress made in the research of PEHs in SEHS for SHM applica-tions, several limitations remain to be addressed, such as how to extract sensing information using the output of a PEH. Since the output generated by a PEH under structural vibration is related to the state of the structure, sensing can be performed directly through the analysis of power amplitude data. Cahill et al. identified the occurrence and growth of cracks in the beam by monitoring the changes in the output power values of the piezoelectric device [22]. Also, the dynamic characteristics information contained in the voltage signal has gradually attracted attention, which is conducive to more detailed structural state assessment and damage identification. For example, it is feasible to extract the structural frequency from the output signal of the piezoelectric device on the bridge pier to infer the bridge scour defect [23]. However, the sensing method that only relies on monitoring power values has low sensitivity and is susceptible to environmental interference. So it is difficult to distinguish structural behaviour changes caused by noise in complex environments from signals indicating minor damage, which raises concerns about the accuracy of such sensing [28].

Recently, machine learning (ML) algorithms have become popular for enhancing the adaptability and sensitivity of SHM through continuous training and learning [29]. With advancements in artificial intelligence, ML has matured significantly in detecting structural damage based on vibration and acceleration data, such as identifying bridge cracks [30] and deck deterioration [31]. However, there is still a gap in employing machine learning to perform damage feature extraction of bridges in PEH-based SEHS systems. Nguyen et al. achieved accurate recognition of structural anomalies in steel railway bridges under train passages by training Convolutional Neural Network (CNN) [32]. In addition, the effectiveness of ML algorithms in SEHS applications of piezoelectric devices is also demonstrated in the supervised framework proposed by Peralta et al. [27].

Nevertheless, in such supervised ML approaches, the cost and availability of labelled data limit the practicality of the research, particularly as the inaccessibility of damage data poses a significant obstacle to implementing bridge SHM. Unsupervised learning, such as convolutional variational autoencoders (CVAE), do not require labelling, which is the key to improving the robustness of bridge damage identification [33]. Therefore, this study extends previous work [30] to develop an unsupervised learning framework for bridge SHM that achieves effective energy harvesting while enabling accurate damage identification, utilising solely the voltage signals from healthy bridge conditions.

The main contributions of this paper are outlined as follows:

- We demonstrate that a piezo-electric energy harvester (PEH) attached to the bridge, can act as a sensor for bridge damage detection, while producing useful electric energy from the bridge vibration.

- We propose a sensing framework for bridge damage detection based on an unsupervised learning algorithm, trained only on voltage signals produced by a PEH on the healthy bridge.

- By employing a vehicle bridge interaction model and experimental database, we investigate how the performance of a PEH is affected by complex environmental factors, including varying levels of sensing noise.

- According to different damage scenarios including various severity, damage location

and sensing location, we perform bi-objective optimisation of a PEH design and summarise design guidelines for optimal performance. This not only improves the overall performance of piezoelectric devices, but also expands their applicability in complex environments.

The remainder of this work is organised as follows. The details of the numerical models in the simulation framework are introduced in Section 2. Section 3 presents the simulation parameters for the case of piezoelectric devices and structures, as well as the experimental setup for the lab-scale VBI system. Section 5 shows the results of applying the framework introduced in Section 4, including the acceleration-based baseline case and the voltage-based study case. Conclusions are given in Section 6, along with suggestions for future works.

## 2. Simultaneous Energy Harvesting and Sensing

In order to explore the performance of piezoelectric devices in SEHS, this section presents a comprehensive numerical platform including bridge acceleration inputs, energy harvesting of PEHs and an unsupervised damage detection method. The first module is the vehicle bridge interaction model, which simulates the vibration response of the bridge under vehicle operation to provide flexible input data for the platform based on multiple damage states. Using acceleration signals as inputs, PEHs are then modelled to enable energy harvesting and voltage signal generation. Finally, a sensing module integrating signal processing and an unsupervised algorithm is proposed to identify the damage state of the bridge. By combining these components, the energy harvesting and sensing capabilities of a PEH device for bridge SHM are effectively defined.

### 2.1. Vehicle Bridge Interaction Model

A VBI model to simulate the dynamic response of the bridge under different working conditions is presented, as shown in Fig.1. By applying a representative half-car model traversing a simply supported bridge, the dynamic coupling equations effectively characterise the responses of both undamaged and damaged bridges [34].

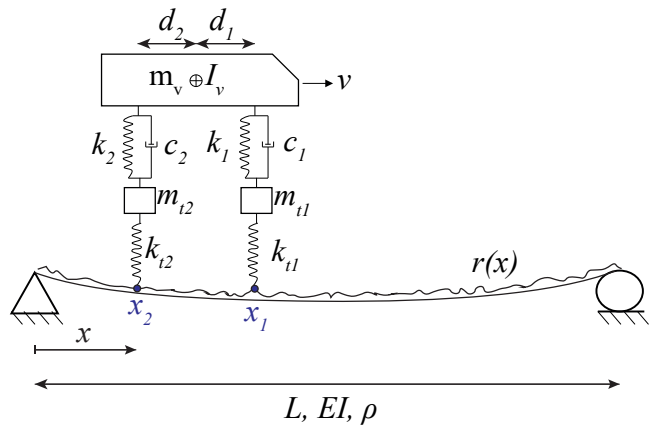

Fig. 1: Illustration of VBI for a half-car vehicle model travelling over a simply supported beam

*2.1.1. VBI Model*

In the vehicle bridge interaction simulation of this study, the bridge model is based on a simply supported Euler-Bernoulli beam, and the equation of motion, discretised with finite element method, is described in Eq. 1.

$$\mathbf{M}_b\ddot{\mathbf{Z}}_b(t) + \mathbf{C}_b\dot{\mathbf{Z}}_b(t) + \mathbf{K}_b\mathbf{Z}_b(t) = \mathbf{F}_b(t), \tag{1}$$

where $\mathbf{Z}_b(t)$ is the displacement vector of the bridge, $\mathbf{M}_b$, $\mathbf{C}_b$, and $\mathbf{K}_b$ are the bridge mass, damping, and stiffness matrices, respectively, and $\mathbf{F}_b(t)$ is the force vector due to the interaction with the vehicle. Full expressions for these matrices and vectors are given in [35].

The vehicle is modelled as a two-axle system travelling at constant speed $v$ with four DOFs: the vertical displacement of the body $z_c(t)$, the rotation of the body $\theta_c(t)$, and the vertical displacements of the front and rear tires $z_{t1}(t)$ and $z_{t2}(t)$ respectively. The equation of motion for the vehicle is given as

$$\mathbf{M}_v\ddot{\mathbf{z}}_v(t) + \mathbf{C}_v\dot{\mathbf{z}}_v(t) + \mathbf{K}_v\mathbf{z}_v(t) = \mathbf{F}_v(t), \tag{2}$$

where $\mathbf{z}_v(t) = [z_c(t), \theta_c(t), z_{t1}(t), z_{t2}(t)]^T$ is the vector of vehicle generalized coordinates, $\mathbf{M}_v$, $\mathbf{C}_v$, and $\mathbf{K}_v$ are the mass, damping, and stiffness matrices of the vehicle system, and $\mathbf{F}_v(t) = [F_{c1}(t), F_{c2}(t)]^T$ is the force vector due to the interaction with the bridge, where $F_{c1}(t)$ and $F_{c2}(t)$ are the contact forces at each axle described as follows

$$\mathbf{F}_v(t) = \begin{bmatrix} F_{c1}(t) \\ F_{c2}(t) \end{bmatrix} = \begin{bmatrix} k_{t1}\left(z_{t1}(t) - z_b(x_1, t) - r(x_1)\right) \\ k_{t2}\left(z_{t2}(t) - z_b(x_2, t) - r(x_2)\right) \end{bmatrix} \tag{3}$$

where $z_b(x_1, t)$ and $z_b(x_2, t)$, are the bridge responses at the location of the front and rear axles, respectively, at time $t$, and $r(x_1)$ and $r(x_2)$ are the road profiles heights at the locations of the the front and rear axles, respectively. $k_{t1}$ and $k_{t2}$ are the stiffness of the front and rear tires, respectively.

The vehicle mass, damping, and stiffness matrices are as follows

$$\mathbf{M}_v = \begin{bmatrix} m_v & 0 & 0 & 0 \\ 0 & I_v & 0 & 0 \\ 0 & 0 & m_{t1} & 0 \\ 0 & 0 & 0 & m_{t2} \end{bmatrix}, \quad \mathbf{C}_v = \begin{bmatrix} c_1 + c_2 & c_1d_1 - c_2d_2 & -c_1 & -c_2 \\ c_1d_1 - c_2d_2 & c_1d_1^2 + c_2d_2^2 & -c_1d_1 & c_2d_2 \\ -c_1 & -c_1d_1 & c_1 & 0 \\ -c_2 & c_2d_2 & 0 & c_2 \end{bmatrix},$$

$$\mathbf{K}_v = \begin{bmatrix} k_1 + k_2 & k_1d_1 - k_2d_2 & -k_1 & -k_2 \\ k_1d_1 - k_2d_2 & k_1d_1^2 + k_2d_2^2 & -k_1d_1 & k_2d_2 \\ -k_1 & -k_1d_1 & k_1 + k_{t1} & 0 \\ -k_2 & k_2d_2 & 0 & k_2 + k_{t2} \end{bmatrix} \tag{4}$$

where, considering $i = 1, 2$ representing the front and rear axles respectively, $m_{ti}$ refers to the mass of the tires, $m_v$ is the vehicle body mass, $I_v$ is the vehicle's rotational moment of inertia, $c_i$ is the damping of the suspension, $d_i$ is the distance from the center of mass to the axles, $k_i$ is the stiffness of the suspension, and $k_{ti}$ is the stiffness of the tire.

The solution to the VBI system is obtained iteratively, using Newmark-$\beta$ method for time integration, with time step $\Delta t = 0.001$ s and 100 elements in the FEM discretisation. Further details on the formulation of the VBI model can be found in [36].

*2.1.2. Road Roughness Profile*

The road surface roughness $r(x)$ in the above governing equations, is described by a power spectral density (PSD) function $G_d(n)$ based on ISO 8608 [37].

$$G_d(n) = G_d(n_0)(n/n_0)^{-w} \tag{5}$$

where the spatial frequency per meter is denoted by $n$, with the parameters $w = 2$, $n_0 = 0.1$ cycle/m, $\Delta n = 0.04$ cycle/m. By defining the road profile amplitude $\eta$ as $\sqrt{2G_d(n)\Delta n}$, the road roughness profile is given as follows.

$$r(x) = \sum \eta_i \cos(n_i x + \theta_i) \tag{6}$$

where $\eta_i$, $n_i$ and $\theta_i$ represent the $i$th amplitude, spatial frequency and random phase angle, respectively. Note that the roughness type $G_d(n_0)$ consists of multiple sets of values defined in ISO 8608 [37]. Specifically, this study adopts road type A with a value of $16 \times 10^{-6}$ m$^3$ and road type B with a value of $64 \times 10^{-6}$ m$^3$.

*2.1.3. Damaged Bridge Model*

The identification of different bridge damage states requires exploring damage models of bridge structures. In this work, we use a simple beam crack model [38], which assumes a linear reduction in bridge stiffness in the interval 1.5 times the beam height on the both sides of the crack. Although this model is only applicable to a single plane as it is assumed that cracks do not cause other internal cracks, it is acceptable to simulate the response of damaged bridges under vehicle operation. The flexural stiffness of the damaged bridge is as follows,

$$EI_e(\zeta) = \begin{cases} EI_0 - E(I_0 - I_c)\dfrac{\zeta - \zeta_1}{\zeta_c - \zeta_1} & \text{if } \zeta_1 \leq \zeta \leq \zeta_c \\ EI_0 - E(I_0 - I_c)\dfrac{\zeta_2 - \zeta}{\zeta_2 - \zeta_c} & \text{if } \zeta_c \leq \zeta \leq \zeta_2 \end{cases} \tag{7}$$

where the parameters of the beam are defined as Young's modulus $E$, width $b$, height $h$ and crack depth $h_c$. Additionally undamaged and cracked second moment of inertia are denoted by $I_0$ and $I_c = b(h - h_c)^3/12$, respectively. The spatial coordinate and the position of the crack are represented as $\zeta$ and $\zeta_c$, respectively. Parameters $\zeta_1$ and $\zeta_2$ are the position $1.5h$ away from the left and right sides of the crack, respectively. Note, that according to this model, the crack is parameterised with two parameters: severity $h_c$ and location $\zeta_c$.

*2.2. PEH Model*

In this work, PEH is modelled as a bimorph configuration, with a three-layer cantilevered plate structure, as shown in Fig.2. It consists of two layers of piezoelectric ceramic materials connected in series to an external electrical resistance, with a non-piezoelectric structural layer in the middle to enhance stability. Note, that piezo-electric material partially covers the substructure layer. The geometry of the device is designed as a rectangle with five parameters: the length $L$ and the width $W$ of the PEH, the length $L_{pzt}$ and thickness $h_p$ of the piezoelectric material, and the thickness $h_s$ of the substructure. The total width and thickness of the PEH device are given by $W = W_{pzt} = W_s$, $h = 2h_p + h_s$, respectively.

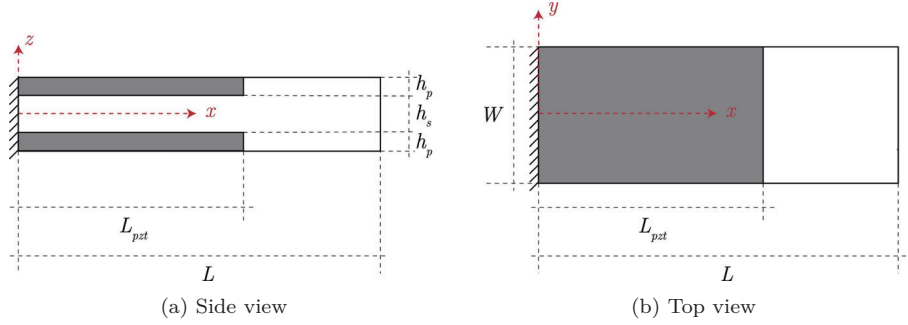

Fig. 2: Schematic of the bimorph PEH [39]

The PEH model based on Kirchhoff-Love plate theory and Hamilton's principle developed by Peralta et al. [40] is applied in this study. Isogeometric analysis (IGA) with B-Splines is selected as the numerical method to discretise the resulting weak form of the Kirchhoff-Love plate problem due to its higher computational efficiency in comparison with standard finite element analysis [39]. It is necessary to explain that the modelling of PEH is based on the following assumptions: all layers in PEH including piezoelectric layers and non-piezoelectric structure layers are perfectly bonded; the piezoelectric material is covered with conductive and negligible thickness electrodes. Then, based on the Kirchhoff-Love plate theory, the PEH model is assumed to only consider in-plane strains rather than transverse shear strains, i.e., ignoring the rotational degrees of freedoms, so the displacement field $\boldsymbol{u}$ and the mechanical strain components $\mathbf{S}$ in terms of the vertical displacement ($w$) of the mid-plane at time $t$ are described by Eq.8 and Eq.9. Note that the coordinate system $(x, y)$ lies on the neutral plane ($z = 0$).

$$\boldsymbol{u} = \{u, v, w\}^T = \left\{-z\frac{\partial w}{\partial x}, -z\frac{\partial w}{\partial y}, w\right\}^T \tag{8}$$

$$\mathbf{S} = \{\varepsilon_x, \varepsilon_y, 2\varepsilon_{xy}\}^T = \left\{-z\frac{\partial^2 w}{\partial x^2}, -z\frac{\partial^2 w}{\partial y^2}, -2z\frac{\partial^2 w}{\partial x \partial y}\right\}^T \tag{9}$$

The constitutive equations for the piezoelectric layer under the plane-stress are as follows,

$$\begin{bmatrix} \mathbf{T} \\ \mathbf{D} \end{bmatrix} = \begin{bmatrix} \mathbf{c}_p^E & -\mathbf{e}^T \\ \mathbf{e} & \varepsilon^S \end{bmatrix} \begin{bmatrix} \mathbf{S} \\ \mathbf{E} \end{bmatrix} \tag{10}$$

where the vector of electric displacement and electric field components are defined by $\mathbf{D}$ and $\mathbf{E}$, respectively. Based on the plane stress assumption, the elastic stiffness matrix $\mathbf{c}_p^E$, permittivity components matrix $\varepsilon^S$, and piezoelectric constant matrix $\mathbf{e}$ are composed of material parameters as follows,

$$\mathbf{c}_p^E = \begin{bmatrix} \bar{c}_{11}^E & \bar{c}_{11}^E & 0 \\ \bar{c}_{12}^E & \bar{c}_{22}^E & 0 \\ 0 & 0 & \bar{c}_{66}^E \end{bmatrix}, \quad \mathbf{e} = \{\bar{e}_{31}, \bar{e}_{32}, 0\}^T, \quad \boldsymbol{\varepsilon}^S = \bar{\varepsilon}_{33}^S \tag{11}$$

Based on IGA, the shape $\mathbf{x}$ in parametric space $\xi \in [0, 1] \times [0, 1]$ and the displacement $w = w(\xi, t)$ are given as linear expansions with B-Spline basis $N_I(\xi)$, control points $\tilde{\mathbf{x}}_\mathbf{I}$ and control variables $w_I(t)$, where $I$ is a bivariate index and $N$ is the number of B-Spline basis functions, as

$$\mathbf{x}(\xi) = \sum_{I=1}^{N} N_I(\xi)\tilde{\mathbf{x}}_\mathbf{I} \tag{12}$$

$$w(\xi, t) = \sum_{I=1}^{N} N_I(\xi)w_I(t) \tag{13}$$

The governing electro-mechanical equations of the PEH are as follows,

$$\mathbf{M}\ddot{\mathbf{w}} + \mathbf{C}\dot{\mathbf{w}} + \mathbf{K}\mathbf{w} - \mathbf{\Theta}v_p = \mathbf{F}a_b \tag{14}$$

$$C_p\dot{v}_p + \frac{v_p}{R_l} + \mathbf{\Theta}^\mathbf{T}\dot{\mathbf{w}} = 0 \tag{15}$$

where $\mathbf{w} \in \mathbb{R}^{N \times 1}$, $\mathbf{M} \in \mathbb{R}^{N \times N}$, $\mathbf{K} \in \mathbb{R}^{N \times N}$, $\mathbf{C} = \alpha\mathbf{M} + \beta\mathbf{K} \in \mathbb{R}^{N \times N}$, $\mathbf{F} \in \mathbb{R}^{N \times 1}$ and $\mathbf{\Theta} \in \mathbb{R}^{N \times 1}$ denote the vector of control variables $w_I(t)$, the mass matrix, the stiffness matrix, the mechanical damping matrix, the mechanical forces vector, the electromechanical coupling vector, respectively. Full expressions of all matrices are given in [40]. $\alpha$ and $\beta$ represent the proportional damping coefficients. $a_b$ defines the input acceleration, and $v_p$ represents the output voltage. $C_p$ and $R_l$ are the capacitance and the optimised external electric resistance, respectively. In a particular case of harmonic excitation, according to the relationship between the output voltage ($v_p = V_o e^{i\omega t}$) and the input acceleration ($a_b = A_b e^{i\omega t}$), the voltage frequency response function (FRF) is defined as follows ($i = \sqrt{-1}$).

$$H_v(\omega) = \frac{V_o(\omega)}{A_b(\omega)} = i\omega\left(\frac{1}{R_l} + i\omega C_p\right)^{-1}\mathbf{\Theta}^T\left(-\omega^2\mathbf{M} + i\omega\mathbf{C} + \mathbf{K} + i\omega\left(\frac{1}{R_l} + i\omega C_p\right)^{-1}\mathbf{\Theta}\mathbf{\Theta}^T\right)^{-1}\mathbf{F} \tag{16}$$

Next, the modal order reduction is performed to reduce the computational burden under long-term signals and optimisation [39], considering only a limited number of modes ($K$) within a certain frequency range. The reduced electro-mechanical equations of PEH are,

$$\ddot{\boldsymbol{\eta}} + \mathbf{c}_o\dot{\boldsymbol{\eta}} + \mathbf{k}_o\boldsymbol{\eta} - \boldsymbol{\theta}_o v(t) = \mathbf{f}_o a_b(t) \tag{17}$$

$$C_p\dot{v}(t) + \frac{v(t)}{R_l} + \mathbf{\Theta}^T\mathbf{\Phi}_o\dot{\boldsymbol{\eta}} = 0 \tag{18}$$

where the matrix $\mathbf{\Phi}_o \in \mathbb{R}^{N \times K}$, $\mathbf{c}_o \in \mathbb{R}^{K \times K}$ and $\mathbf{k}_o \in \mathbb{R}^{K \times K}$ are reduced mode shape matrix containing only the first $K$ mode shape vectors $\boldsymbol{\phi}_i$, reduced mechanical damping and the reduced stiffness, respectively. Also, $\boldsymbol{\theta}_o$, $\boldsymbol{\eta}$ and $\mathbf{f}_o \in \mathbb{R}^{K \times 1}$ are the electro-mechanical coupling vector, modal coordinates and mechanical forces vector, respectively. The PEH model is written as

$$\dot{\mathbf{Z}} = \mathbf{Q} \cdot \mathbf{Z} + \mathbf{b} \cdot a_b(t), \text{ where } \mathbf{Z} = \begin{bmatrix} \boldsymbol{\eta} \\ \dot{\boldsymbol{\eta}} \\ v_p \end{bmatrix}, \quad \mathbf{Q} = \begin{bmatrix} 0 & 1 & 0 \\ -\mathbf{k}_o & -\mathbf{c}_o & \boldsymbol{\theta}_o \\ 0 & -\frac{\mathbf{\Theta}^T\mathbf{\Phi}_o}{C_p} & -\frac{1}{C_p R_l} \end{bmatrix}, \mathbf{b} = \begin{bmatrix} 0 \\ \mathbf{f}_o \\ 0 \end{bmatrix} \tag{19}$$

In order to obtain the voltage signal for a given input acceleration signal, the system of differential equations (Eq.19) is solved using the MATLAB Simulink ode45 solver. In addition, the amount of energy harvested by the piezoelectric device in time interval $[t_1, t_2]$ is defined as

$$E = \int_{t_1}^{t_2} \frac{v^2(t)}{R_l} dt \tag{20}$$

Therefore, the PEH model mentioned above allows the generation of voltage time histories and energy output based on inputs of excitation signals beyond just harmonic vibrations. Note, that this model was developed in [39] and [40], where it was verified against finite element method and validated against experimental data. In this study, the PEH model is used to estimate the voltage signal produced from the bridge vibration. Two study cases are considered: bridge vibration simulated by the VBI model and bridge vibration measured in the lab experiement.

*2.3. Damage Detection*

To define the sensing accuracy for bridge damage detection, the components in the sensing framework, including signal processing and a deep learning algorithm, are described below.

*2.3.1. Wavelet Synchrosqueezed Transform*

As bridge damage detection relies on identifying changes in the structure's fundamental modal information [30], a clear and precise frequency representation and analysis are crucial. However, due to noise contamination caused by environmental disturbances, it is difficult for traditional Fourier analysis to accurately capture the local features in the signal. It is beneficial to employ time-frequency representation under local transformation to provide more detailed information for the sensing process, such as Continuous Wavelet Transform (CWT). Under the wavelet $\psi$, the CWT of signal $x(t)$ at scale $a$ and $b$ is represented by $W_x$ [41].

$$W_x(a, b) = \int x(t) a^{-1/2} \psi\left(\frac{t-b}{a}\right) dt \tag{21}$$

As an adaptation of the CWT, the wavelet synchrosqueezed transform (WSST) improves the resolution by introducing a squeezing operation [42]. The WSST ($S_x$) with threshold $\gamma$ is defined as follows [41],

$$S_x^\gamma(\omega, b) = \int_{|W_x(a,b)|>\gamma} W_x(a, b) \delta(\omega - \hat{\omega}_x(a, b)) \frac{da}{a} \tag{22}$$

where $\hat{\omega}$ and $\delta$ denote the instantaneous frequencies and the Dirac distribution, respectively.

Therefore, in this study, WSST image is utilised to transform input data such as acceleration or voltage signals into corresponding high-resolution time-frequency spectra. This processing approach not only provides a clear visualisation of time-frequency characteristics but also effectively enhances the characterisation of damage features within the signals.

Furthermore, detecting minor frequency changes due to slight damage is challenging, e.g., a damage severity of 10% results in only a 0.98% shift, so Hurtado et al. [33] enhanced the visualisation by narrowing the frequency range (i.e., the y-axis of the image). Similarly, in the present study, the frequency range is chosen to include the first two fundamental frequencies of the bridge. This range not only provides the most critical insights for damage detection but also contributes to the high output of PEHs under the low-frequency operation of the bridge, as confirmed by Yao et al. [14], which is sufficient for SEHS applications.

*2.3.2. Convolutional Variational Autoencoder*

As mentioned before, to evaluate PEH's performance in sensing applications, an autoencoders (AE) is used to learn meaningful features from input data. This method enables accurate identification of structural damage characteristics by analysing the voltage response generated during energy harvesting. Being an unsupervised methodology, it is trained with healthy structure samples.

An AE is composed of an encoder and a decoder for feature learning and extraction, as shown in Fig.3, where $x = [x_1, x_2, \ldots, x_n]$ is the unlabelled input data, which is dimensionally reduced to the latent space $y = [y_1, y_2, \ldots, y_p]$ via the encoder. Then, the decoder reconstructs the input as $x' = [x'_1, x'_2, \ldots, x'_n]$ from the latent vector. The weight matrices $\mathbf{W}$ and $\mathbf{W}'$ are used to connect neurons, and the transformations between layers are as follows,

$$y = \mathrm{f}\left(\mathbf{W}x + b\right) \tag{23}$$

$$x' = \mathrm{g}\left(\mathbf{W}'y + b'\right) \tag{24}$$

where f and g are activation functions of the encoder and decoder; bias vectors in decoding and encoding are represented by $b$ and $b'$. The feature learning of AE is achieved by minimising the reconstruction error (Eq.25).

$$Loss\left(x, x'\right) = \|x - x'\|^2 \tag{25}$$

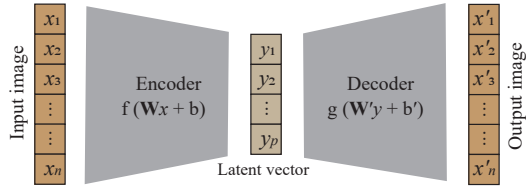

Fig. 3: Illustration of the AE structure

However, a significant drawback of AE is that the distribution of the latent space lacks continuity [43]. Variational Autoencoders (VAEs) introduce regularization in the latent space to better capture the underlying structure of the data. Latent variables $y$ are sampled from a probability density function $p(y)$ defined over a high-dimensional space $\mathcal{Y}$. Deterministic functions $f(y;\theta)$, parametrised by $\theta \in \Theta$, map elements from $\mathcal{Y}$ to the input data space $\mathcal{X}$, expressed as $f : \mathcal{Y} \times \Theta \to \mathcal{X}$. The goal is to optimise $\theta$ such that $f(y;\theta)$, with $y \sim p(y)$, approximates the input dataset $x$, as:

$$p(x) = \int p(x|y;\theta)p(y)dy \tag{26}$$

In the training process of the VAE, the encoder maps the input $x$ to a Gaussian distribution characterized by its mean $\mu$ and covariance matrix $\sum_x$, such that $y \sim \mathcal{N}(\mu, \sum_x)$. This Gaussian distribution is the prior encoding distribution $q(y|x)$. The sampling step is performed using the *re-parametrisation trick*, defined as: $y = \mu + \sigma\epsilon$, where $\sigma$ represents the standard deviation, and $\epsilon \sim \mathcal{N}(0,1)$. This step ensures that the latent space remains differentiable, enabling back-propagation during training [44]. The decoder then learns a set of parameters $\theta$ to approximate $p(x|y;\theta)$, reconstructing the input $x$ based on samples from the prior distribution [44].

The optimisation of the VAE is performed by maximizing the Evidence Lower Bound (ELBO) loss, which combines two components: the reconstruction loss $\mathcal{L}_1$ and the Kullback-Leibler (KL) divergence $\mathcal{L}_2$. The reconstruction loss $\mathcal{L}_1$ measures the difference between the input $x$ and the reconstructed output $x'$, while the KL divergence $\mathcal{L}_2$ quantifies the discrepancy between two probability distributions. The ELBO loss is given as:

$$\text{ELBO} = \mathcal{L}_1(x, x') + \mathcal{L}_2(q(y|x)||p(y)), \tag{27}$$

with the respective components defined as:

$$\mathcal{L}_1(x, x') = (x - x')^2, \tag{28}$$

$$\mathcal{L}_2(q(y|x)||p(y)) = \int q(y|x) \ln \frac{q(y|x)}{p(y)} \, dy. \tag{29}$$

In practice, the distributions $q(y|x)$ and $p(x|y)$ are approximated using fully connected neural networks in VAE. However, this work employs a variant of VAE called the Convolutional Variational Autoencoder (CVAE), which replaces the fully connected layers with convolutional layers from CNNs. CNNs are more effective at extracting features from image data, such as the time-frequency domain representations of voltage signals used in this study [30]. Fig.4 depicts the structure of the CVAE employed in this study. Further details on the architecture and hyper-parameter selection of the CVAE used in this work can be found in [30]. Note that the training parameters for the CVAE are set to use a batch size of 32, trained for 100 epochs, and a learning rate of $10^{-3}$.

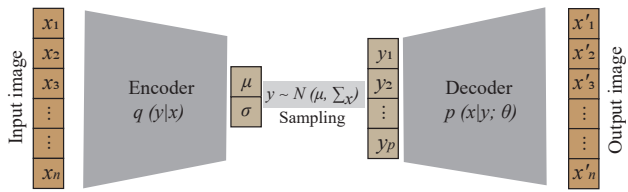

Fig. 4: Illustration of the CVAE structure

*2.3.3. Sensing Accuracy*

This section defines sensing accuracy to evaluate and quantify the sensing capabilities of PEHs. Sensing accuracy is defined with respect to how many samples in a set of healthy and damaged samples are identified correctly by a trained CVAE.

As described in Section 2.3.1, it is feasible to adopt WSST images as input and output of CVAE to achieve sharper visualisation and enhance the characterisation of feature content in the signal. In order to quantify the error between samples, the Mean Square Error based on the original $(S_i)$ and reconstructed image $(\bar{S}_i)$ represents the Damage Index $(DI)$ of the $i$-th sample, which is defined as follows.

$$DI_i = \left(S_i - \bar{S}_i\right)^2 \tag{30}$$

Based on the $DI$ value, the sensing accuracy $S$ is defined by the percentage of correctly classified samples, as shown below. Let $[DI_1, DI_2, \ldots, DI_n]$ represent the DI values for $n$ samples. In order to indicate whether the $i$-th sample is correctly classified, a correctness indicator $\delta_i \in \{0, 1\}$ is introduced, where 1 and 0 denote correct and incorrect classifications, respectively.

$$\delta_i = \begin{cases} 1, & \text{if } (DI_i \leq \Gamma \text{ and } s_{h,i} = 1) \vee (DI_i > \Gamma \text{ and } s_{d,i} = 1), \\ 0, & \text{otherwise.} \end{cases} \tag{31}$$

where $\Gamma$ is the damage threshold; $s_{h,i} = 1$ or $s_{d,i} = 1$ indicates that the original state of the $i$-th sample is healthy or damaged, respectively. Then

$$S = \frac{1}{n} \sum_{i=1}^{n} \delta_i \tag{32}$$

Note that the selection of a practical threshold depends on the operational characteristics of each specific structure. In this section, the threshold is defined as the 90th percentile based on previous work [30] and similar report [45].

## 3. Case Studies

In this work, we focus on the simultaneous optimisation of a PEH for energy harvesting and sensing under bridge acceleration inputs. The optimisation begins with a single design variable - device length, $L$, to better explore the objective functions and the relationship between the two objectives. In section 5.3, the study is extended to a design space with two design variables - $L$ and the aspect ratio $R = W/L$. It is worth noting that the framework can be further extended to more complex design spaces in a straightforward way.

The parameters of a PEH are given in Table 1, where the thickness ratio is defined as $H = h_p/h$, the length ratio as $\ell = L_{pzt}/L$. The design variable $L$ is assumed to be in the range $[0.1 - 0.5]$m. The selection of such a range is based on our previous studies and it aims to optimise the coupling between the devices' natural frequencies and the first two vibration modes of the bridge, which are significant for vibration energy harvesting [14].

Table 1: General characteristics of the PEH

| Property | Value | Property | Value |
|---|---|---|---|
| Device length, $L$ [m] | [0.15 - 0.5] | Length ratio, $\ell$ | 0.1 |
| Device width, $W$ [cm] | $R \times L$ | Aspect ratio, $R$ | 1 |
| Device thickness, $h$ [m] | 0.01 | Thickness ratio, $H$ | 0.3 |
| Young's Modulus, $E$ [GPa] | 105 | $c_{11}^E, c_{22}^E$ [GPa] | [69.5, 69.5] |
| $c_{12}^E$ [GPa] | 24.3 | $c_{66}^E$ [GPa] | 22.6 |
| $\epsilon_{33}$ [nF/m] | 9.57 | $d_{31}^E, d_{32}^E$ [C/m$^2$] | [-16, -16] |
| Density, $\rho_s$ [kg/m$^3$] | 9000 | PZT Density, $\rho_p$ [kg/m$^3$] | 7800 |
| Poisson's ratio, $\nu$ | 0.30 | PZT Poisson's ratio, $\nu$ | 0.30 |
| Damping, $\alpha$ [rad/s] | 14.65 | Damping, $\beta$ [s/rad] | $1 \times 10^{-5}$ |

*3.1. Numerical Case Study*

In order to reveal the capabilities of PEHs in the field of damage detection and explore the effects of multiple variables including damage severity, damage location, installation location of PEHs and road roughness, the VBI model, described in Section 2.1, is used in the first numerical study. Parameters of the bridge and the two-axle vehicles are given in Table 2. In order to capture more realistic dynamic response characteristics, the randomness of the database is ensured by randomly selecting four vehicle parameters (mass, speed, distance between axles, and relative position of the front axle with respect to the centre of mass) from predefined intervals in previous studies [30]. These parameters are sampled using the Latin Hypercube Sampling (LHS) method, following a uniform distribution within the specified interval. Consequently, each acceleration signal is generated based on a unique set of randomly sampled vehicle parameters, ensuring variability across the dataset.

For the bridge model defined in Table 2, the modal frequencies ($f_n$) of the healthy bridge is obtained according to $\frac{n^2\pi}{2L_b^2}\sqrt{\frac{EI_0}{\mu}}$, with the first two fundamental frequencies being 4.8 Hz ($f_1$) and 19.1 Hz ($f_2$), respectively. Therefore, all signal images used for training and validation are narrowed to the frequency range of [0, 20] Hz.

Table 2: Parameters of the VBI model

| Property | Value | Property | Value |
|---|---|---|---|
| Span length, $L_b$ [m] | 25 | Cross section area, $A$ [m$^2$] | 8.7 |
| Second moment of inertia, $I_o$ [m$^4$] | 2.9 | Young's Modulus, $E$ [GPa] | 2.87 |
| Mass per bridge length, $\mu$ [kg/m] | 2303 | Bridge damping ratio, $\xi$ | 3% |
| Vehicle mass, $m_v$ [t] | [0.5 - 1.5] | Vehicle speed, $v$ [km/h] | [50 - 60] |
| Distance between axles, $d_1, d_2$ [m] | [2 - 3.5] | Front axle position from centre, $x$ | [0.4 - 0.5] |
| Axle suspension stiffness, $k_1, k_2$ [N/m] | 27500 | Axle suspension damping, $c_1, c_2$ [Ns/m] | 1300 |
| Axle tire stiffness, $k_{t_1}, k_{t_2}$ [N/m] | $1.5 \times 10^5$ | Axle tire damping, $c_{t_1}, c_{t_2}$ [Ns/m] | 5 |

Several bridge states are investigated in this work; healthy state (HN), damage at midspan (DMN) and damage at the quarter-span (DQN). Also, the damage severity is represented by the ratio of crack depth to bridge beam height based on the crack model described in Section 2.1.3. We use two damage severities corresponding to crack ratios of 10% and 20%, denoted by '1' and '2', respectively, amended to the damage state notation. For example, DMN1 denotes the numerical state corresponding to 10% damage severity at the mid-span. Table 3 indicates the shift of the first two fundamental frequencies of the bridge under different damage scenarios, which is the key to identifying damage.

Table 3: The first two bending modes of the bridge for multiple states

| Bending Modes | HN | DMN1 | DMN2 | DQN1 |
|---|---|---|---|---|
| Mode 1, $f_1$ [Hz] | 4.8 | 4.6 (4.2%) | 4.4 (8.3%) | 4.7 (2.1%) |
| Mode 2, $f_2$ [Hz] | 19.1 | 19.0 (0.5%) | 19.0 (0.5%) | 18.5 (3.1%) |

*3.2. Experimental Case Study*

In order to further investigate practical application of the proposed SEHS optimisation framework, the second case study includes a lab-scale VBI set up, shown in Fig.5. The main bridge structure consists of a steel I-section beam. As shown in Fig.6, the beam is simply supported at both ends and has a span of 10.6m. More details are provided in [46]. It is worth noting that vehicle movement is guided by square steel rails mounted on the bridge deck surface, as illustrated in Fig.5.

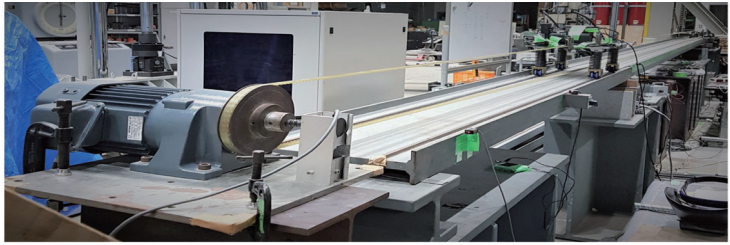

Fig. 5: A lab-scale VBI set up: simply supported beam represents the bridge [46]

To simulate damage, additional mass is added at a specific location on the bridge. This produces a reduction in the natural frequency of the bridge, similar to the one caused by bridge cracking [47]. Fig.7(a) illustrates the damage scenario implemented in this study, where a 16.1 kg mass is placed at mid-span (DME). A scaled two-axle vehicle, shown in Fig.7(b), incorporats a power unit and an electronic controller to set a constant velocity. The scaled vehicle has a total weight of 20.2 kg and an axle spacing of 44 mm.

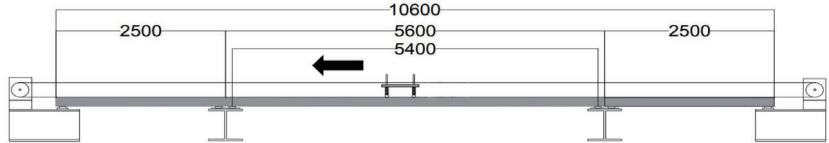

Fig. 6: Schematic of the beam (dimensions in millimeters) [46]

For data acquisition, TML ARS-A accelerometer with a sampling frequency of 200Hz is installed at mid-span of the bridge for continuous signal monitoring. The experimental data in this study corresponds to vehicle passage at a constant speed of 0.04 m/s.

To approximate the effects of environmental disturbances present in real bridges, random hits at varying locations and intensities are applied to the beam, artificially introducing perturbations in the output signals. Furthermore, the designed surface roughness (class A) of the vehicle's tracks also contributes to random variations in the response data under vehicle-bridge interaction. This roughness profile follows the same specifications used in the numerical studies of this work, in accordance with ISO 8608 [37], and is derived from a real bridge with 40.4 m span [48].

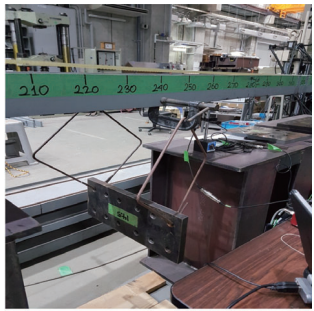 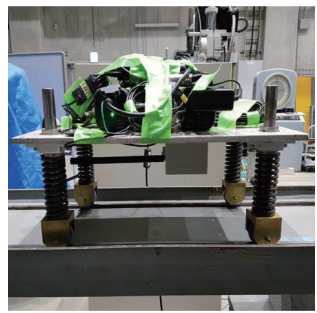

(a) Concentrated mass  (b) A scaled two-axle vehicle

Fig. 7: A lab-scale VBI set up: bridge damage, represented by a concentrated mass at mid-span, and the vehicle [46]

## 4. Optimisation Framework of SEHS

In this section, an unsupervised SEHS framework is proposed to design and optimise PEHs for simultaneous energy harvesting and damage detection. The framework consists of four phases, as shown schematically in Fig.8 and Fig.9. The first three phases show evaluation of sensing accuracy and energy harvesting performance of a PEH with fixed design $L$. The fourth phase represents the optimisation process to find the optimal designs of PEH (Fig.9), i.e., device that maximises both - sensing accuracy and energy production. The four phases are explained in more details below.

The purpose of **Phase 1** is to evaluate the amount of energy harvested from 500 vehicle passes over a healthy bridge and 100 passes over a damaged bridge. In each pass, vehicle parameters are chosen randomly from the distributions specified in Table 2 for the VBI study and a scaled two-axle toy car for the experimental case study. The study is repeated for 36 PEHs with lengths $L = [0.15, 0.16, \ldots, 0.5]$ m. For each device and a given bridge state (healthy or damaged), an average energy over all vehicle passes is recorded and denoted as $E(L)$.

In **Phase 2**, the WSST algorithm is employed as the signal processing tool to convert all voltage signals, generated in **Phase 1**, into visualised time-frequency domain images. These images serve as the input for CVAE.

In **Phase 3**, the unsupervised algorithm based on CVAE, as described in Section 2.3.2, is applied to access the sensing accuracy of each PEH. WSST images converted from 500 voltage signals under the healthy bridge, are split with an 80:20 ratio for training and validation purposes, respectively. Then, 100 healthy and 100 damaged samples serve as the testing set to define the sensing accuracy in Eq.32, denoted as $S(L)$.

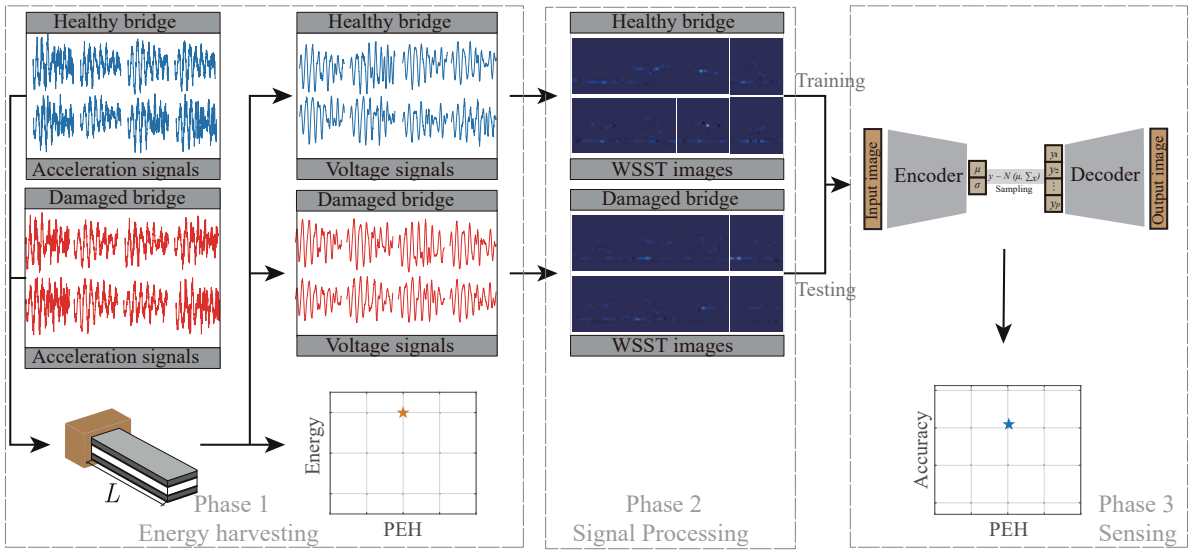

Fig. 8: SEHS framework

Finally, in **Phase 4**, a bi-objective optimisation algorithm is employed to investigate the relationship between the two objectives (energy production and sensing accuracy) across various damage scenarios, (Fig.9). The goal is to identify a series of optimal Pareto front points, each representing a specific design (length $L$) of the PEH. The optimisation problem is formalised as follows,

$$L^* = \arg \max_{L \in [0.15, 0.5]} \{E(L), S(L)\} \tag{33}$$

However, a significant difficulty brought by the optimisation process is the heavy computational load caused by estimating the harvested energy and sensing accuracy for each PEH within a large design interval. This arises from the multiple steps involved in SEHS, including database generation and processing, as shown in Fig.8. To ease the computational burden, the Kriging surrogate model [49] is employed to approximate $E(L)$ and $S(L)$.

Regarding the optimisation process, the Non-Dominated Sorting Genetic Algorithm II (NSGA-II) is an ideal candidate for solving the bi-objective problem in this work. This is because it ensures diversity and efficiency in exploring the Pareto front solution by the incorporation of crowding distance and elitist strategy [50]. The resulting Pareto front provides valuable insights into the design of PEHs in SEHS and offers a range of optimised solutions for various operational scenarios.

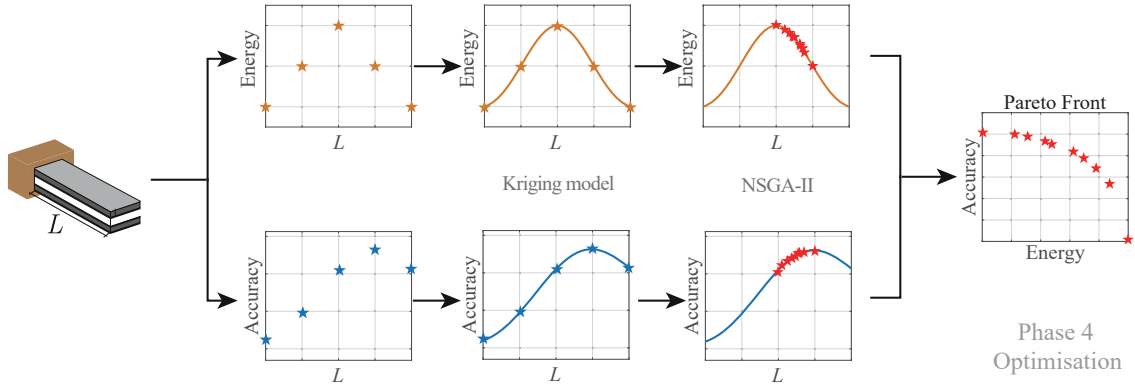

Fig. 9: Bi-objective PEH design optimisation framework: first energy and sensing accuracy are evaluated for a discrete set of parameter $L$, then functions $E(L)$ and $S(L)$ are approximated for all values of $L$ using Krigging model, then NSGA-II is employed to find Pareto front of optimal solutions

## 5. Results and Discussion

The following results show the performance of the proposed framework in PEH-based SHM under the numerical and experimental case studies, introduced in Section 3. Note that in the damage detection process, in addition to conducting voltage-based sensing, acceleration-based sensing is used as a benchmark. Acceleration-based damage detection is performed by using the bridge acceleration signal directly as input for CVAE and evaluating the corresponding accuracy, as described in Section 2.3.

*5.1. Numerical Case Study: Optimisation of PEHs under DMN1*

First, the unsupervised SEHS framework (Fig.8) is implemented and verified based on the acceleration signal and the voltage signal recorded during the energy harvesting process when a 10% crack occurs at the mid-span of the bridge structure (DMN1) under road type A.

*5.1.1. Acceleration-based Sensing*

In this section, the results of sensing based on acceleration signals in the bridge are presented. As mentioned previously, 500 healthy and 100 damaged acceleration signals are simulated through the VBI model. Fig.10(a, b) show examples of acceleration signals and the corresponding WSST images, respectively, for a healthy (HM) and damaged (DMN1) bridges.

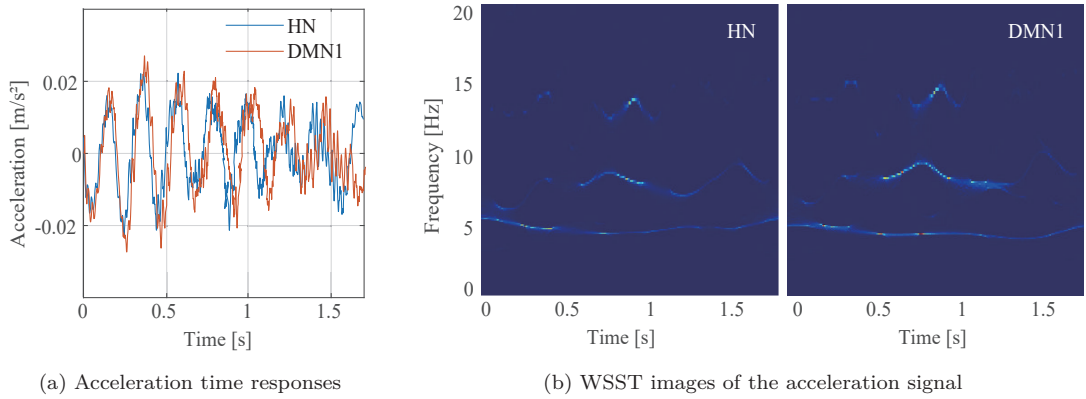

(a) Acceleration time responses  (b) WSST images of the acceleration signal

Fig. 10: Examples of the acceleration signals and WSST images under HN and DMN1 bridge states

For each input acceleration WSST image, the output of the CVAE model is a damage index, DI. Distribution of DI values for a healthy and damaged bridge are shown in Fig.11(a), where it can be observed that two distributions have an overlap, meaning that the classification accuracy cannot reach 100% and will depend on the chosen threshold. For 90% threshold, the confusion matrix is shown in Fig.11(b), where the green and red backgrounds represent the number of correctly classified healthy and damaged samples, respectively. The average acceleration-based sensing accuracy for DMN1 is 81%, indicating the feasibility of the unsupervised approach for damage identification.

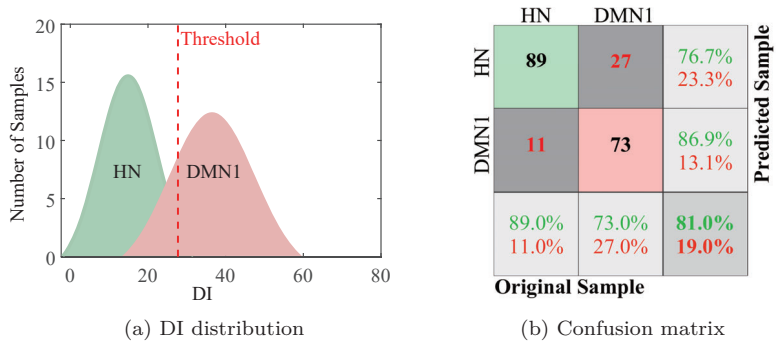

(a) DI distribution  (b) Confusion matrix

Fig. 11: Results of the acceleration-based damage detection for DMN1

*5.1.2. Voltage-based Sensing*

In order to illustrate the dependence of the sensing accuracy of the PEH on its design, we first consider three cases corresponding to $L = 0.15, 0.25$ and $0.35$m. Other material and geometry parameters are given in Table 1. Damage scenario DMN1 (10% damage at mid-span) is considered. PEH is placed at mid-span. The Frequency Response Functions (FRFs) of the PEHs with different lengths, together with the FFT of a sample acceleration signal, are shown in Fig.12, where first two frequencies of the bridge $f_1$ and $f_2$ are also indicated. It can be observed that with changes in the device length $L$, the FRFs of the three PEHs exhibit significant differences, which may have a notable effect on sensing accuracy. Note, that the fundamental frequency of the device with $L = 0.35$m is close to the fundamental frequency of the bridge, which is most affected by the damage located at the mid-span. However, from this observation alone we cannot conclude that this device will be the best for damage identification. To further illustrate the difference between the three devices in terms of the outcome voltage signals and their corresponding WSST images, in Fig.13, voltage signals generated from the same vehicle passage over a healthy and damaged bridge (i.e., from the same acceleration signal) are shown.

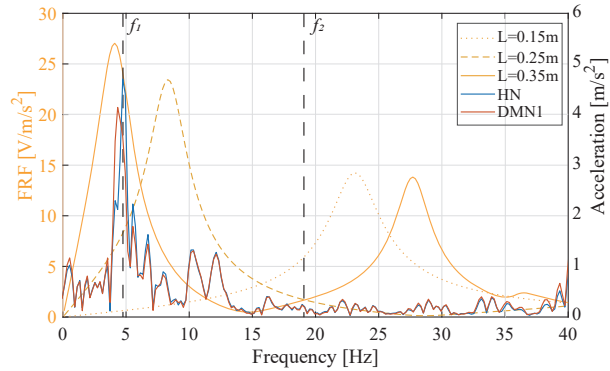

Fig. 12: FRFs of three PEHs together with the FFT of an acceleration signal

In Fig.13, it can be seen that the voltage signals from three PEHs exhibit differences in the WSST images. This is because some PEHs respond not only to the first vibration mode but also to other frequency components (e.g., responses around 10 Hz), leading to variations in voltage signals. The time signals in Fig.13 also reveal the potential of PEHs as filters, showing varying capabilities in noise elimination due to the diversity of their natural frequencies, which leads to the necessity to employ a rigorous sensing framework.

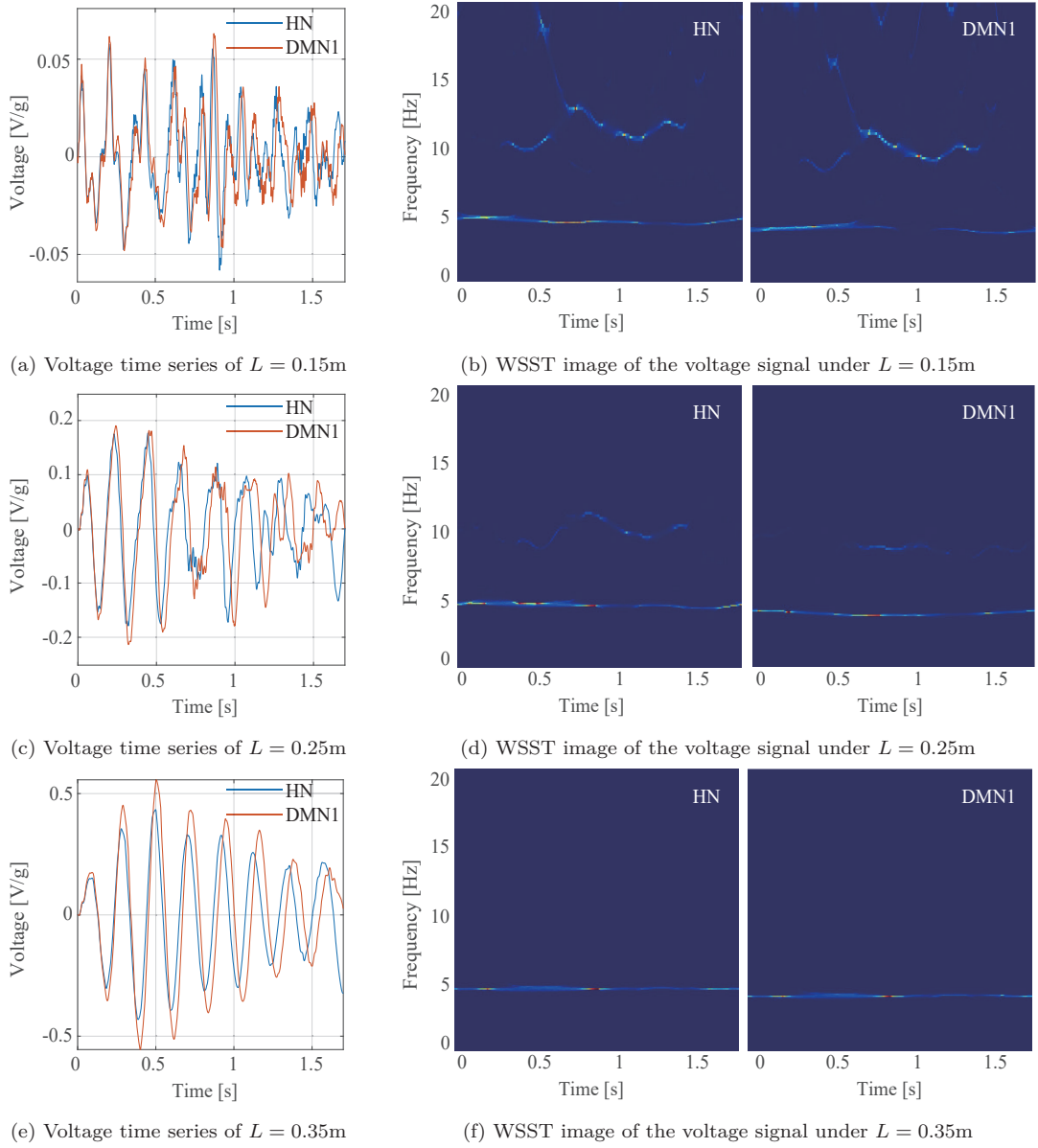

(a) Voltage time series of $L = 0.15$m  
(b) WSST image of the voltage signal under $L = 0.15$m  
(c) Voltage time series of $L = 0.25$m  
(d) WSST image of the voltage signal under $L = 0.25$m  
(e) Voltage time series of $L = 0.35$m  
(f) WSST image of the voltage signal under $L = 0.35$m  

Fig. 13: Illustration of the voltage signals of three PEHs with $L = 0.15$, $0.25$ and $0.35$m

Fig.14 presents the sensing results based on voltage signals and CVAE. After testing five times, the average value and stability (indicated by the error bar) of the accuracy from the three devices and the benchmark based on acceleration signal are shown in Fig.14. As seen in Fig.12, as $L$ decreases, the resonance frequency of the PEH deviates further from the first natural frequency of the bridge ($f_1$). The dominance of the first fundamental frequency of PEHs enhances the electromechanical response under other vibration components, such as signals that do not contain damage information as shown in Fig.13(b, d), which are regarded as noise components in the unsupervised sensing process, resulting in a loss of accuracy. And different levels of interference cause the diversity of the DI distribution and sensing accuracy under different lengths of PEH. For instance, the PEH with a length of 0.15m exhibits the worst accuracy, as it contains stronger noise interference. On the other hand, the PEH with a length of 0.35m, which is close to the first fundamental frequency of the bridge, demonstrates the best sensing performance, achieving an 11% improvement and excellent stability in accuracy compared to traditional acceleration signals.

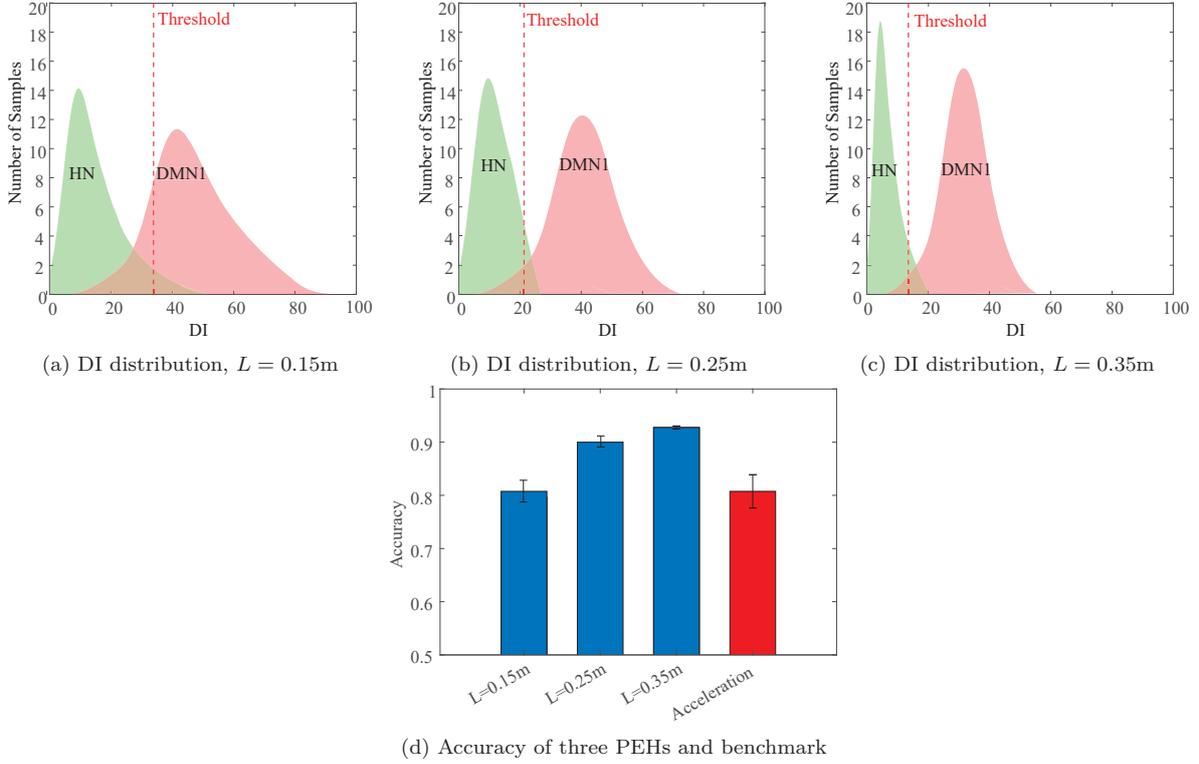

Fig. 14: Voltage-based sensing for three various designs of PEHs: distribution of DI indices for healthy and damaged bridges, and overall accuracy in comparison with acceleration-based sensing

The above results demonstrate the feasibility of PEH serving as a sensor in bridge damage detection and the effectiveness of the proposed unsupervised framework that only requires healthy state data. It is also shown that the sensing accuracy varies with the length of the device, and reasonable design is able to reduce the impact of interference signals. The inability to distinguish meaningful changes in the voltage signal for certain PEH configurations highlights the importance of optimising the PEH design to ensure accurate and reliable damage detection. Therefore, the design optimisation of PEH with length as variable is then performed based on the framework proposed in Section 4.

*5.1.3. Bi-objective Optimisation of SEHS*

This section shows the application of the optimisation framework in Fig.9 under DMN1. As described in **Phase 4** of Section 4, Kriging models based on $E(L)$ and $S(L)$ are trained to reduce the computational burden. The Kriging approximation of energy output and sensing accuracy within the PEH design space is presented in Fig.15, demonstrating an excellent fit between the predicted points and the original (support) points. The subscript denotes the damage scenario, such as $E_{\text{HN}}$ representing the amount of energy harvested by each device from the healthy state of the bridge. In Fig.15(a), more energy is harvested under the damaged state because the loss of bridge stability produces more available mechanical vibrations. The results in Fig.15(b) show that the accuracy of PEHs in detecting 10% damage is generally superior to acceleration-based sensing due to the potential filtering capability.

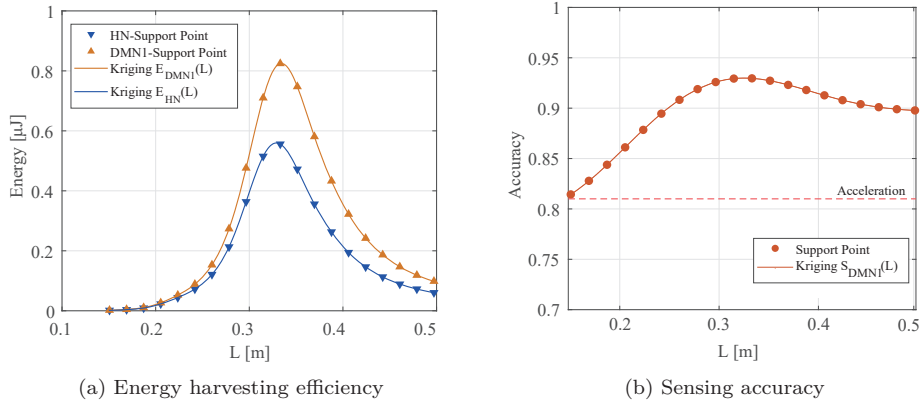

(a) Energy harvesting efficiency  (b) Sensing accuracy

Fig. 15: Kriging approximations of the energy and sensing under healthy and DMN1 bridge conditions

Also, as seen from Fig.15, both, energy production and sensing accuracy are dependent on $L$. Fig.16 illustrates the optimal candidates (Pareto front) represented by stars obtained by NSGA-II. It is obvious that the framework proposed in this study is able to identify ideal PEHs under the joint consideration of two objectives. From Fig.16, it can be seen that the Pareto front under the current scenario tends to show the correlation between the two objectives, and it is possible to identify a device that maximises both objectives simultaneously: $L = 0.34$m. This device has a fundamental frequency of 4.8Hz, which matches the fundamental frequency of the bridge. This is because in this damage scenario, the resonance effect results in highest energy harvesting and highest amplification of the shift in the first frequency due to damage, hence highest damage identification accuracy. Consequently, the best performance in both objectives is achieved simultaneously. To further confirm this result, we conducted two more studies with damage at the mid-span. In the first study, the damage severity was increased to 20% (DMN2) and in the second study, the PEH was located at the quarter-span (while damage of 10% was kept at the mid-span, i.e. DMN1). Detailed results of these two cases are given in the Appendix, and show similar conclusions as above, i.e. bi-objective optimisation identified a single optimal device - a PEH with $L = 0.34$ m and the fundamental frequency matching the fundamental frequency of the bridge. Note, that in all the aforementioned studies, road type A was used. This road roughness introduced only a small amount of noise into the VBI model, which did not interfere with the resonance effect. For larger noise levels, this interference can be more significant, potentially revealing additional frequencies with high energy harvesting potential. This aspect is explored in the next section.

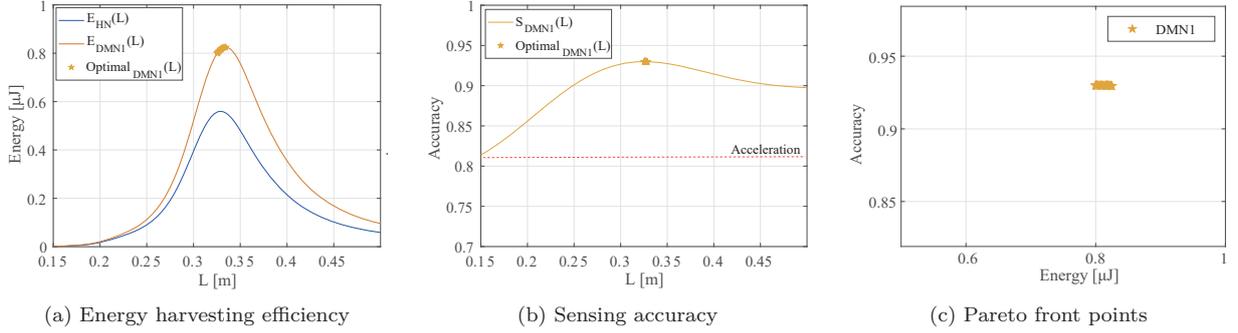

(a) Energy harvesting efficiency  (b) Sensing accuracy  (c) Pareto front points

Fig. 16: Results of bi-objective PEH optimisation under DMN1

*5.1.4. Numerical Case Study: Impact of Road Roughness*

As mentioned above, components of the voltage signal that do not contain damage information act as noise interference, which leads to a reduction in the sensing efficiency of a PEH. Therefore, in order to improve the stability of piezoelectric devices as sensors, it is necessary to conduct studies on sensing accuracy under different noise interference. To study the impact of road noise on the sensing performance of PEHs, three road types containing different roughness are investigated. Based on the road surface roughness $r(x)$ in the VBI model described in Eq.6, the roughness type is denoted by a different value of $G_d(n_0)$ according to the ISO 8608 [37]. In addition to Road A ($16 \times 10^{-6}$ m$^3$), Road B with a value of $64 \times 10^{-6}$ m$^3$ is used to show the diversity of roughness in this section. The smooth pavement without roughness (NR), i.e. $r(x) = 0$, is also considered for the comparison study.

The acceleration signals recorded on three types of road surface roughness under 10% damage at the mid-span are presented in Fig.17, which explains the degree of interference of different road surface roughness in the bridge response signal. In particular, there is strong interference around the first natural frequency from Road B shown in Fig.17(c). It is important to note that the noise in reality cannot be precisely predicted due to the random nature of the road roughness. Therefore, Fig.17 serves as an exemplary representation, highlighting the potential impact of road surface conditions on the acquired signals.

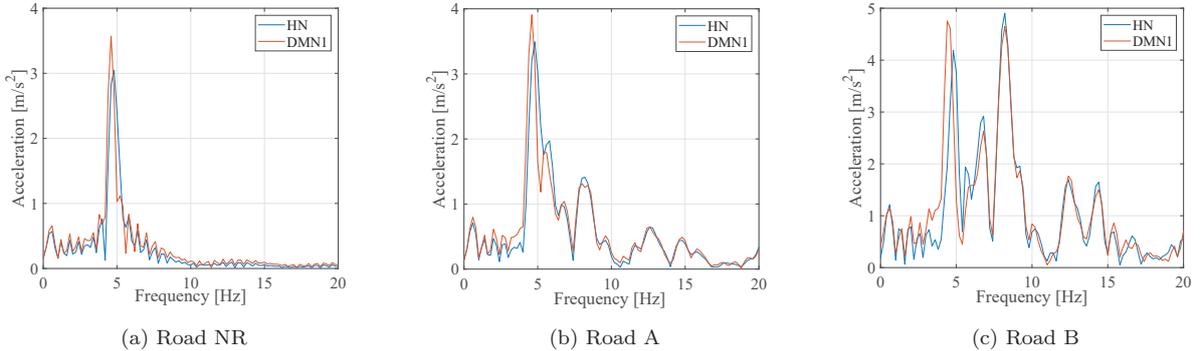

(a) Road NR  (b) Road A  (c) Road B

Fig. 17: Illustration of a Fourier acceleration spectrum for road A, road B and road NR

Fig.18 illustrates the performance of SEHS in enabling PEHs to detect DMN1 across multiple road types. From Fig.18(a), the energy harvested from a road with high surface roughness (type B) is the largest and decreases as the road becomes smoother. This is because smooth roads tend to excite low-amplitude vibrations, which reduces the amount of mechanical energy that can be harvested.

As for the sensing accuracy shown in Fig.18(b), it is evident that under an ideal road surface with no roughness (black line), all PEHs across the entire design space tend to exhibit an excellent ability to distinguish between healthy and damaged states of the structure. This is attributed to the absence of noise interference, which allows all devices to rely solely on the clean first mode frequency to identify any shifts, even for devices with fundamental frequencies substantially different from the target frequency. Moreover, increasing road roughness degrades the accuracy of all PEHs. Additionally, Fig.18(b) highlights varying sensitivity to roughness across PEH designs, with the $L = 0.15$m device experiencing the most severe degradation (a 41% reduction from 95% on smooth roads to 54% on Road B). This significant performance loss demonstrates the challenges of implementing certain PEH designs in practical SEHS applications.

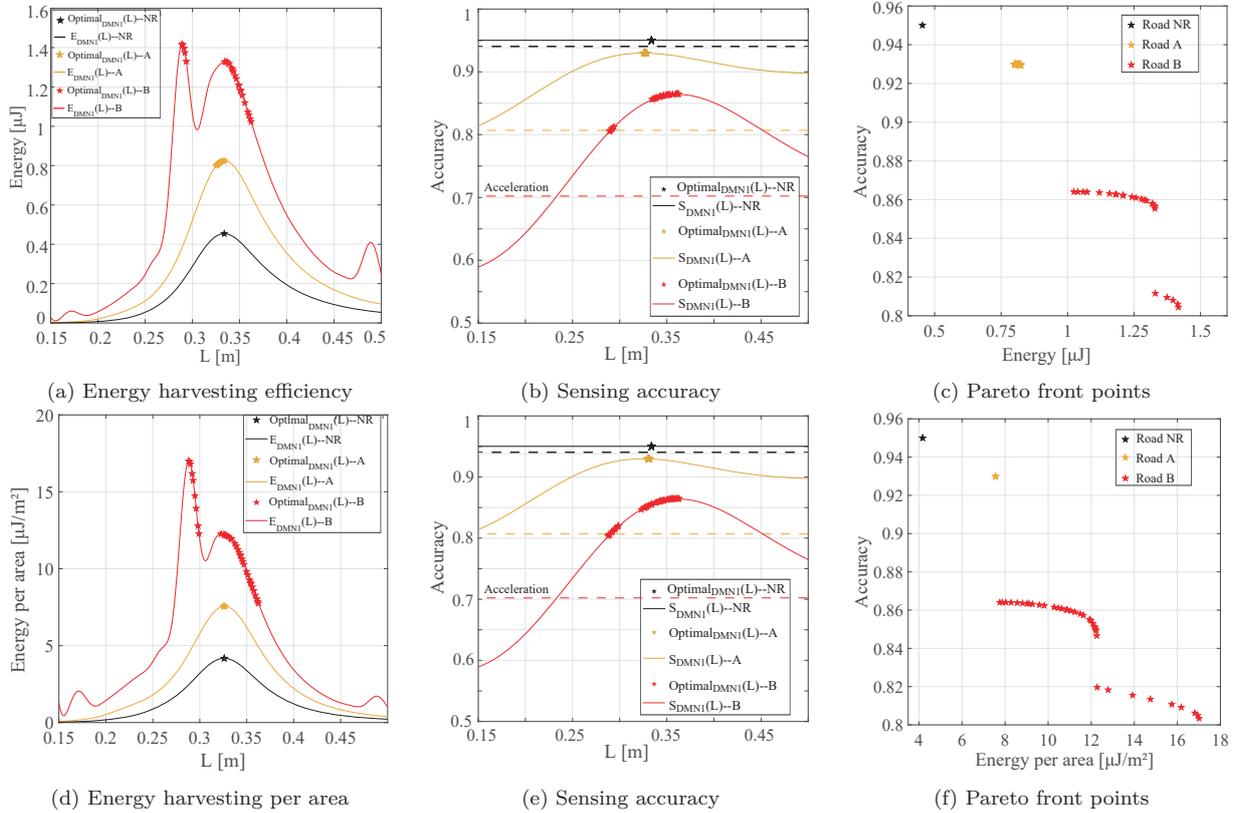

Fig. 18: Results of SEHS under DMN1 for road A, road B and no-roughness (NR). Dashed lines correspond to acceleration-based sensing accuracy

The comparison of three road types cases shows that optimal energy production in the presence of noise due to rough road surfaces increases, from $0.45\mu$J for Road NR to $0.82\mu$J for Road A and $1.33\mu$J for Road B. Notable reduction can be also seen in the accuracy: from 95% for Road NR to 93% for Road A and 86% for Road B. Note that the above comparison is based on the same PEH ($L = 0.34$m). Pareto fronts shown in Fig.18(c) indicate that in the case of Road NR, it is possible to identify a single device, optimal in both objectives, i.e., PEH with length $L = 0.34$m and fundamental frequency of 4.8Hz. The same conclusion was made in Section 5.1.2 for Road A. In the case of Road B, Pareto front demonstrates a significant variation in energy harvesting performance (by 39.2% from $1.02\mu$J to $1.42\mu$J) with a 6% trade-off in the accuracy. Two clusters of Pareto-optimal solutions can be identified, which correspond to accuracies of approximately 86% and 80%, and two peaks of the energy plot in Fig.18(a): devices with $L = 0.29$m and $L = 0.33 - 0.36$m, respectively. While the maximum accuracy of 86% is now reached by device with $L = 0.36$m, this device does not produce maximum energy. Energy production is maximised by the device with $L = 0.29$m. The fundamental frequency of this device is 5.3Hz, which differs from the fundamental frequency of the bridge. This new frequency appeared due to the added noise. This effect is even more pronounced in Fig.18(d, f), where the optimisation is performed in terms of energy per unit area. It can be seen that device with $L = 0.29$m harvests 40% more energy per unit area than device with $L = 0.34$m, while giving up only 3% of sensing accuracy. The overall range of the Pareto front is from $7.9\mu$J/m$^2$ to $17\mu$J/m$^2$ in terms of the energy per unit area and from 80% to 86% in terms of the accuracy. Overall, the above results demonstrate the diversity of Pareto-optimal designs and the trade-off between the two objectives caused by the increase of road roughness, verifying the importance of optimisation for different roads.

Due to the uncertainty of noise in operating environments, the performance of standardised PEH devices becomes difficult to predict and control. Therefore, it is important to conduct sensitivity analysis of a large number of PEHs to find design parameters that are robust to noise. The above results indicate that by implementing the proposed optimisation framework, the design and deployment of PEHs can be customised to improve the stability of PEH-based sensing systems in noisy environments.

*5.2. Numerical Case Study: Impact of Damage Location*

The occurrence of structural damage is usually not limited to a single fixed location, so it is necessary to extend the study to other locations on the bridge. The following study includes the piezoelectric energy harvesting and sensing results based on 10% cracks occurring at one quarter of the bridge (DQN1). It is emphasised that since PEHs are often deployed in a network form, only the devices installed at the same location as the damage occurred are studied here to explore their sensing and optimisation criteria at different bridge locations. Fig.19 is a schematic diagram of a typical acceleration signal in the frequency domain under DQN1. It can be seen that when the damage occurs at the quarter point, the second mode of the bridge is more easily excited, which leads to a obvious shift in the second fundamental frequency of the structure due to damage. Additionally, it can be noticed that due to the damage location, the first natural frequency of the bridge does not vary significantly. The $E(L)$, $S(L)$ and optimisation results obtained during the damage detection process of PEHs under DQN1 are illustrated in Fig.20, which shows a different trend from the DMN1 case.

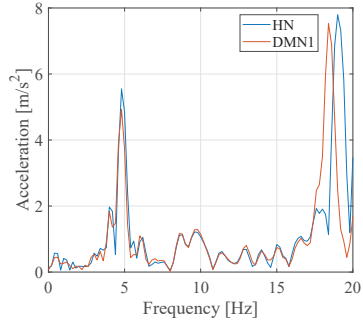

Fig. 19: Illustration of a Fourier acceleration spectrum from DQN1

Regarding energy harvesting, unlike the results under DMN1 (Fig.16(a)), other candidates that are capable of harvesting comparable energy have emerged, besides PEHs having a fundamental frequency tuned to the first frequency of the bridge. As seen from Fig.20, energy plot exhibits three peaks, corresponding to $L = 0.17$m, $0.34$m, and $0.42$m. The FRFs of these three devices and the sample WSST images are shown in Fig.21. It can be seen that device with $L = 0.17$m is tuned to the second frequency of the bridge, and device with $L = 0.42$m has the second frequency equal to the second frequency of the bridge.

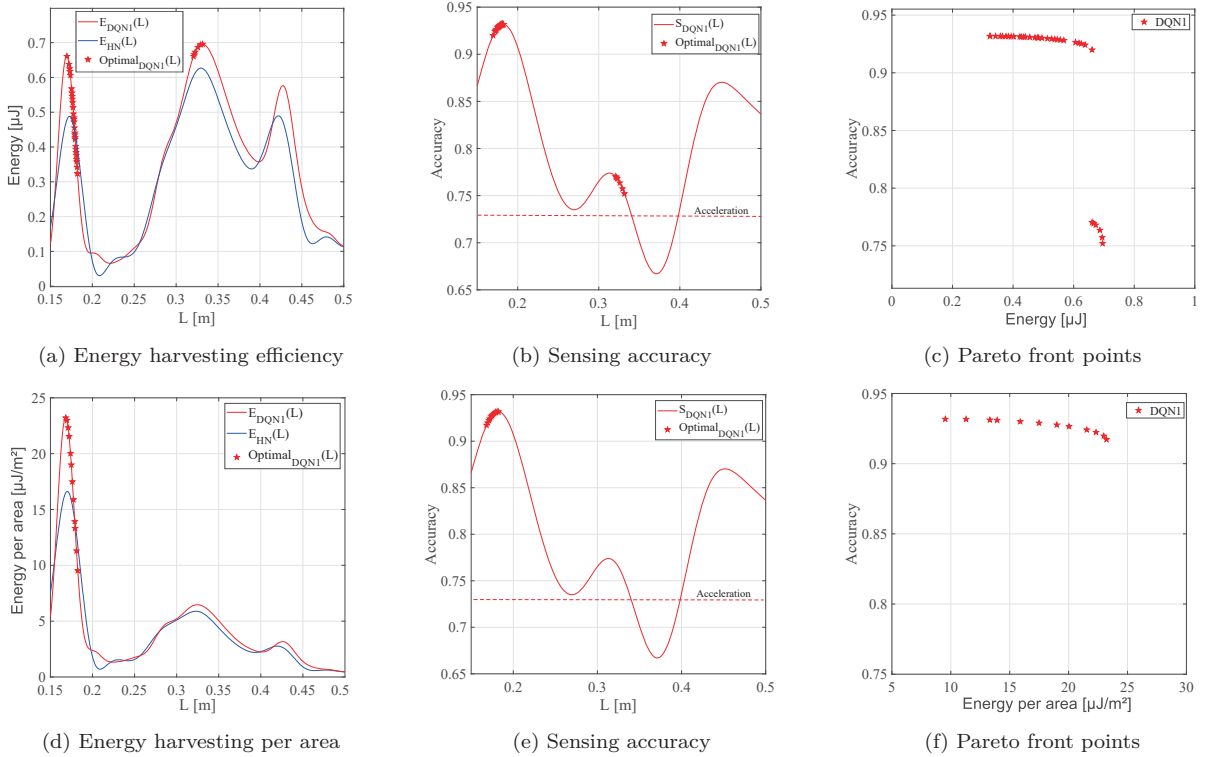

Fig. 20: Bi-objective optimisation results of DQN1

On the other hand, it can be seen from the sensing accuracy depicted in Fig.20(b) that the sensing capability of the piezoelectric device positioned at the quarter point exhibits a different trend in comparison to DMN1. The accuracy of PEH with length of 0.34m drops sharply under the DQN1 scenario, which is mainly attributed to the challenging task of detecting the fundamental frequency shift in the voltage signal, as shown in Fig.21(d). However, due to the high energy output of devices tuned near the first frequency, they are still considered suitable SEHS candidates in Fig.20(c). When considering energy per unit area as the objective (Fig.20(d)), only one optimal cluster is identified. From Fig.20(b, e), it can be concluded that the fundamental frequency of the PEH candidates with the best sensing accuracy is concentrated at the second mode frequency of the structure, which is sufficient to identify the frequency difference between healthy and damaged states (like the PEH in Fig.21(a)). Note, that a case study under Road B is also given in the Appendix.

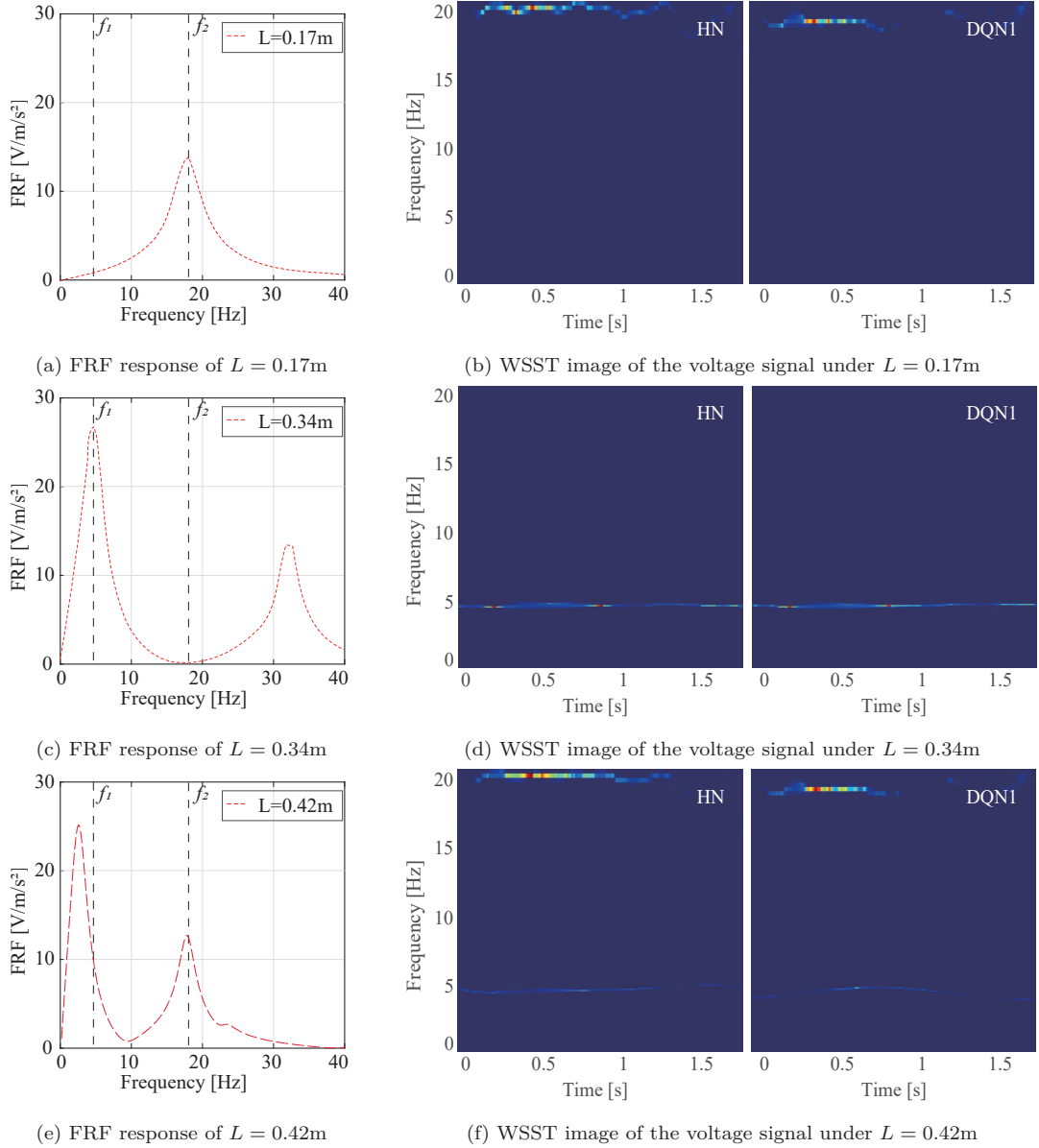

(a) FRF response of $L = 0.17$m  
(b) WSST image of the voltage signal under $L = 0.17$m  
(c) FRF response of $L = 0.34$m  
(d) WSST image of the voltage signal under $L = 0.34$m  
(e) FRF response of $L = 0.42$m  
(f) WSST image of the voltage signal under $L = 0.42$m  

Fig. 21: Illustration of the voltage signals of three PEHs with $L = 0.17, 0.34$ and $0.42$m

By analysing two clusters of solutions in the Pareto front shown in Fig.20(e), we can conclude that two devices, with $L = 0.17$m (i.e. the device tuned to the second frequency of the bridge) and device with $L = 0.34$m (i.e. the device tuned to the first frequency of the bridge), represent two possible optimal solutions. The trade-off between these two designs is: 6.1% increase in terms of energy production (0.66$\mu$J vs. 0.70$\mu$J) and 17.2% decrease in terms of sensing accuracy (93% vs. 77%).

A novel finding is that for devices with other modes that match the target frequency, like the PEH with length of 0.42m (Fig.21(e)), a clear shift of the target frequency can be visualised and provides reliable sensing performance. However, the component of the first mode in the voltage signal has a negative impact on the sensing performance, limiting further improvements in accuracy.

In Fig.21(d, f), the study is repeated in terms of energy per unit area. Here we can clearly observe that the device tuned into the second frequency of the bridge is optimal in both objectives.

*5.3. Case Study of Two Design Parameters*

Studies, presented in the previous sections, highlight the influence of the PEH design on the efficiency of the bridge SEHS, but are limited to a single parameter ($L$). Results indicate a close relationship between natural frequencies of the bridge and those of the PEH. However, in practice, design space can include multiple parameters, and various combinations of design parameters can produce PEHs with the same fundamental frequency. This is illustrated in Fig.22 for a design space consisting of two parameters: length $L$ and aspect ration $R = W/L$, and two options - with and without added tip mass $M_{\text{tip}} = 0.025$kg (denoted as Type A and Type B, respectively).

The PEH fundamental frequency isocurves in Fig.22(a) show that the fundamental frequency is independent of $R$ in the absence of a tip mass. However, in Fig.22(b), in presence of tip mass, the fundamental frequency of the device tends to be affected by both $L$ and $R$, in a non-linear way. The isocurve, corresponding to the bridge frequency $f_1 = 4.8$Hz is highlighted in red, and from the sole analysis of the fundamental frequency, we cannot conclude which device with $w_o = 4.8$Hz will be optimal for both objectives (energy production and sensing accuracy). Therefore, to answer this question, in what follows, we conduct a rigorous optimisation in design space $\{L, R\}$, such that $L \in [0.15, 0.5]$m and $R \in [0.1, 1]$, with a 0.025kg tip mass under Road type A.

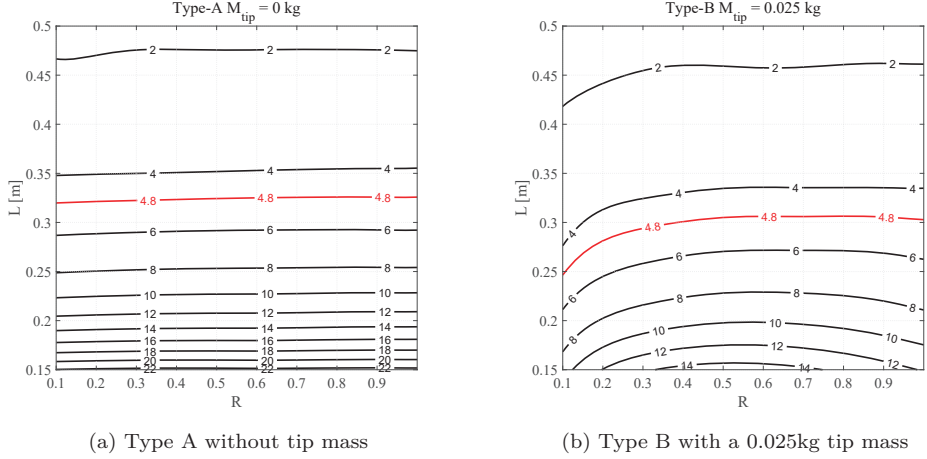

(a) Type A without tip mass  (b) Type B with a 0.025kg tip mass

Fig. 22: Fundamental frequency isocurves for two types of PEHs

Optimisation results under the framework introduced in Section 4 for a 10% damage at the mid-span of the bridge (DMN1) are presented in Fig.23(a, b, c). From Fig.23(a), it can be observed that the ridge of energy function is aligned with the isocurve of 4.8 Hz (as seen in Fig.22(b)), reaching its maximum at point $R = 1$, $L = 0.34$m. Similar behaviour is seen in the accuracy plot, Fig.23(b), where we can additionally observe that all devices with the fundamental frequency of 4.8Hz have the same sensing accuracy. Even though the optimisation algorithm identifies multiple points to form the Pareto front, from Fig.23(c), it can be seen that the difference in accuracy between solutions is negligible and the optimal design can be represented by a single point: $R = 1$, $L = 0.34$m - the far-right point of the 4.8Hz isocurve.

Optimisation results based on energy per unit area are given in Fig.23(d, e, f). In this case, the ridge of energy per unit area function is also aligned with the 4.8Hz isocurve, but now the optimisation algorithm identifies two clusters of solutions: near point $L = 0.25$m, $R = 0.1$, and near point $L = 0.31$m, $R = 0.6$. However, analysing the Pareto front in Fig.23(f), we can see that the difference in accuracy between optimal solutions is negligible, and therefore, device with maximum energy per unit area, i.e. $L = 0.25$m, $R = 0.1$, can be considered optimal for both objectives. It is interesting to notice, that the optimal design in this case corresponds to the far-left point of the 4.8Hz isocurve.

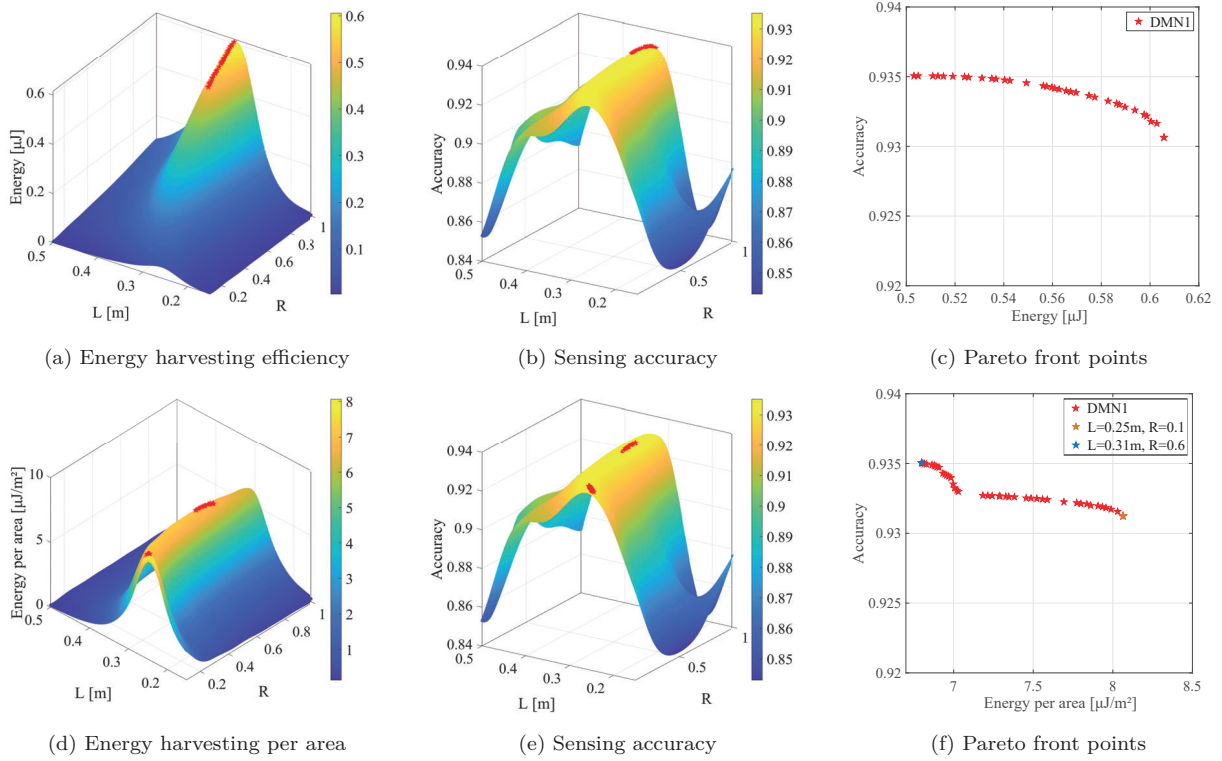

Fig. 23: Results of SEHS under DMN1 for PEHs with a 0.025kg tip mass

*5.4. Experimental Case Study*

This section provides further validation of the practical application of the optimisation framework proposed in Section 4 based on lab-scale data. As mentioned earlier, accelerometers installed at the mid-span of the bridge structure, as described in Section 3.2, are utilised for data acquisition. Therefore, it is assumed that the PEH is placed at the mid-span. The experimental signals are generated by multiple passes of a scaled two-axle vehicle over the beam in both healthy (HE) and damaged (DME) state, where additional mass is suspended at mid-span to represent structural damage.

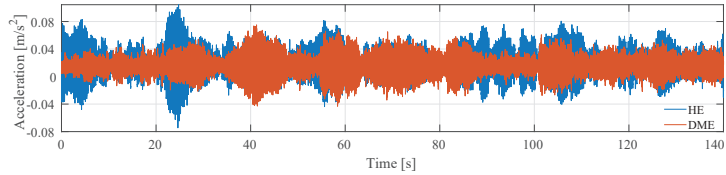

Fig. 24: Illustration of the acceleration signals under HN and DME

Fig.24 illustrates the structural responses in the time domain with 140s. It is noted that the time window for the lab-scale data is longer than that of the numerical simulations, primarily due to differences in vehicle travelling speeds, which may lead to variations in the order of magnitude of energy harvesting outcomes. Additionally, the frequency range of the WSST images is maintained at [0, 20] Hz to ensure the capture of the most significant modal responses at the first fundamental frequency ($f_{lab} = 3.6$ Hz) while maintaining computational efficiency of the training process.

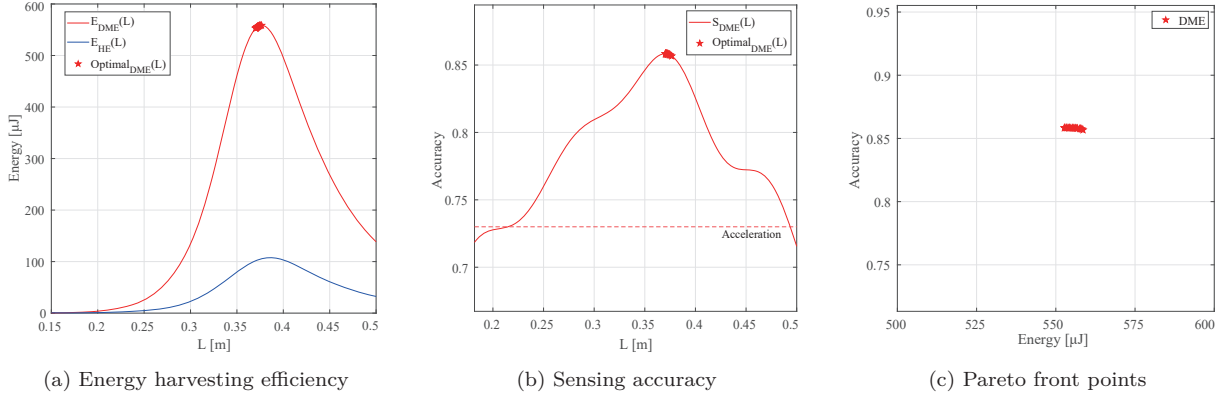

(a) Energy harvesting efficiency    (b) Sensing accuracy    (c) Pareto front points

Fig. 25: Results of bi-objective optimisation under DME

The performance and optimisation results of PEHs in SEHS under the experimental setting are presented in Fig.25. As shown in Fig.25(a, b), the distribution patterns of energy output and sensing accuracy closely resemble those observed in the numerical case where damage occurs at mid-span (Fig.16). Compared to traditional acceleration-based sensing, the piezoelectric devices exhibit superior capability in detecting bridge damage. Notably, the optimised device ($L = 0.37$m in Fig.26) shows clear frequency shift identification in experimental application, achieving a 13% improvement in sensing accuracy compared to accelerometer-based detection. It is again observed, that the same design, tuned to the first frequency of the bridge, is optimal in both objectives - energy harvesting and sensing accuracy - simultaneously.

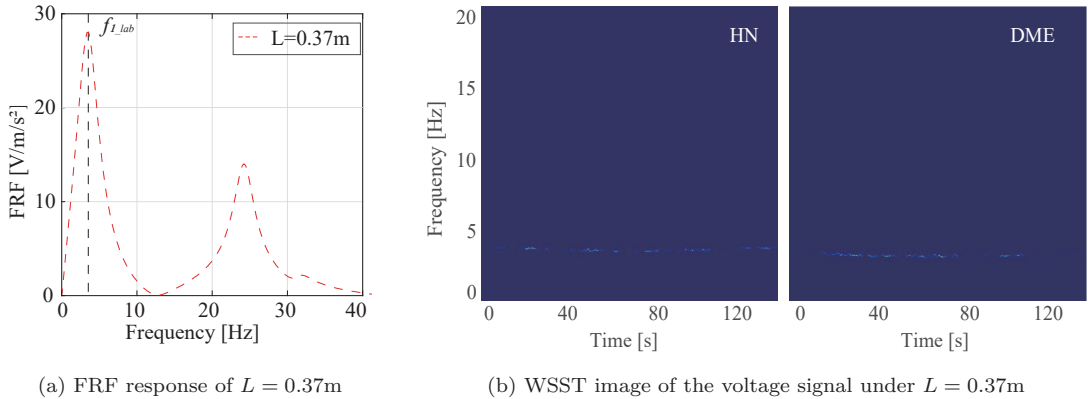

(a) FRF response of $L = 0.37$m    (b) WSST image of the voltage signal under $L = 0.37$m

Fig. 26: Illustration of the voltage signals of the optimal PEH with $L = 0.37$m

In terms of energy efficiency, the comparison between traditional accelerometer-based and the proposed PEH-based SHM is presented in Table 4. As mentioned before, the two sensing components are the TML ARS-A (2V excitation, 120Ω resistance) and the optimal PEH ($L = 0.37$m), with their respective sensing powers denoted as $P_{sensing}$. Also, a low-power

ARM Cortex-M3 MCU is employed for signal acquisition, featuring an average sampling power of $P_{sample} = 480\mu W$ and a sleep power of $P_{sleep} = 6\mu W$ [51]. It is noted that the acquisition time window is set to $t_{sample} = 140s$, where two operating modes are evaluated based on local data storage, including continuous sampling, and sampling with sleep time ($t_{sleep} = 600s$). According to the total energy consumption ($E = [P_{sensing} + P_{sample}] \cdot t_{sample} + P_{sleep} \cdot t_{sleep}$), both modes demonstrate that the PEH-based SHM results in a total energy benefit of approximately 98% compared to the ARS-A.

The result quantifies the significant energy-saving advantage of replacing the conventional sensor with a PEH, highlighting the potential of the proposed optimisation framework to enhance harvesting efficiency and extend the operational lifespan of SHM systems. While other high-demand components of the sensing system, such as acquisition and wireless transmission modules, may still require auxiliary batteries, the substantial reduction in sensing power of the proposed framework lays a strong foundation for achieving fully sustainable operation. It can be further advanced in the future through the development of piezoelectric mata-structure and improved power management strategies.

Table 4: Energy consumption of accelerometer-based and PEH-based SHM under different modes

|  | ARS-A | PEH | ARS-A | PEH |
|---|---|---|---|---|
| Sensing Power, $P_{sensing}[\mu W]$ | $33.3 \times 10^3$ | -4.0 | $33.3 \times 10^3$ | -4.0 |
| Sampling Power, $P_{sample}[\mu W]$ | 480 | 480 | 480 | 480 |
| Sleep Power, $P_{sleep}[\mu W]$ | 6.6 | 6.0 | 6.0 | 6.0 |
| Acquisition Time, $t_{sample}[s]$ | 140 | 140 | 140 | 140 |
| Sleep Time, $t_{sleep}[s]$ | 0 | 0 | 300 | 300 |
| Total consumption, $E[J]$ | 4.729 | 0.067 | 4.733 | 0.068 |

## 6. Conclusion

This work proposed a rigorous design optimisation platform for simultaneous energy harvesting and sensing systems for bridge health monitoring utilising dual-function piezoelectric energy harvesters (PEHs). We showed that PEH can act as a harvester and as a sensor simultaneously, with the sensing task being bridge damage detection. The effectiveness of the CVAE-based unsupervised sensing framework was validated through both simulation and experiment case studies without relying on labelled data. The results highlighted the significant influence of the geometric parameters of PEHs on energy output and sensing accuracy, and NSGA-II was subsequently employed to identify designs optimal in both objectives. We have shown that optimal configurations derived from the proposed framework achieve superior capability to acceleration-based benchmarks in bridge damage detection. In addition, it is shown that the PEH can act as a filter, yielding up to 13% improvement in sensing accuracy of voltage-based sensing while saving 98% of energy consumption in comparison with acceleration-based sensing.

Additionally, to improve the stability and robustness of PEHs as self-powered sensors, we conducted comprehensive bi-objective optimisation studies across multiple scenarios, investigating the effects of pavement roughness, damage severity, damage location, and PEH placement. These findings established clear design guidelines for optimal piezoelectric devices, emphasising the importance of customising the fundamental frequency of PEHs to bridge modal frequencies containing damage information for efficient sensing performance. We have shown that in the absence of noise, there is synergy between the two design objectives, i.e. the same device, tuned to one of the natural frequencies of the bridge, is optimal for both objectives simultaneously. However, in presence of noise, optimal designs form a Pareto front, and the trade-off between two objectives exists.

Future work should explore additional design parameters, including vehicle characteristics and piezoelectric materials, while investigating meta-material designs for more versatile and robust solutions.

## Acknowledgements

This research is undertaken with the assistance of resources and services from the National Computational Infrastructure (NCI), which is supported by the Australian Government. The authors acknowledge support from the University of New South Wales (UNSW) Resource Allocation Scheme managed by Research Technology Services at UNSW Sydney.

The authors also wish to thank Professor C.W. Kim and Professor K.C. Chang, as well as their research teams, for their help in conducting the experiments. Additionally, the author would like to acknowledge the support provided by the Japan Society for the Promotion of Science (JSPS) in conducting this research.

## Appendix:

*Appendix A: Impact of Damage Severity*

First, the previous work (DMN1) is extended to the bridge with 20% damage at mid-span (DMN2). As mentioned before, the damage severity of the bridge in this work is represented by the ratio of crack depth and beam height. Fig.27 shows the energy output and sensing accuracy of PEHs with length as a variable under different damage severity (DMN1 and DMN2). Dashed lines refer to the sensing results under the acceleration benchmark.

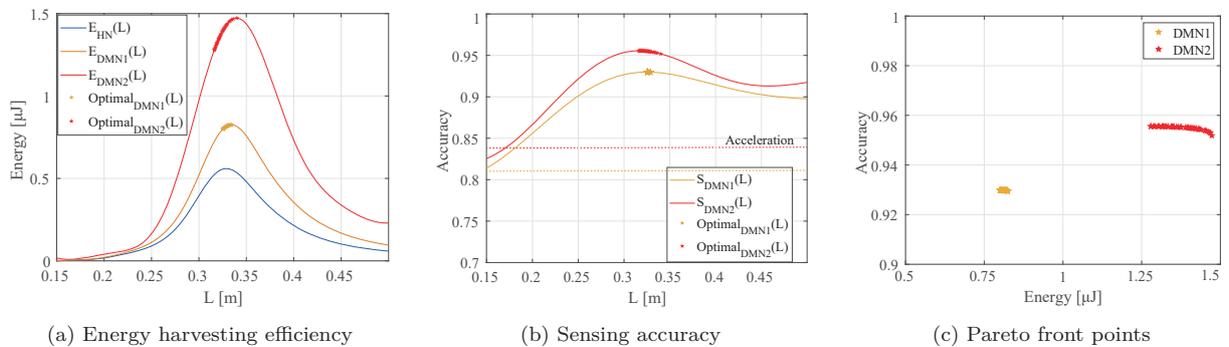

(a) Energy harvesting efficiency     (b) Sensing accuracy     (c) Pareto front points

Fig. 27: Comparative results of DMN1 and DMN2

As shown in Fig.27(a, b), the PEHs under the high damage state DMN2 exhibit higher energy harvesting efficiency and damage identification accuracy. This is based on the fact that higher severity induces further structural instability, resulting in a larger magnitude of vibration excitation and modal frequency shift. In Fig.27(c), it is obvious that PEHs achieve more efficient SEHS performance under high damage severity conditions. However, the optimal configurations of PEHs cluster in similar parameter regions across different damage severities, indicating that damage severity has limited impact on the optimisation results of PEHs.

Another promising finding is that under the unsupervised CVAE algorithm, the distribution of DI at different severities exhibits clear separation, as illustrated in Fig.28. This demonstrates the potential of using the voltage signals from PEH for identifying bridge damage severity, which will be further explored in future research.

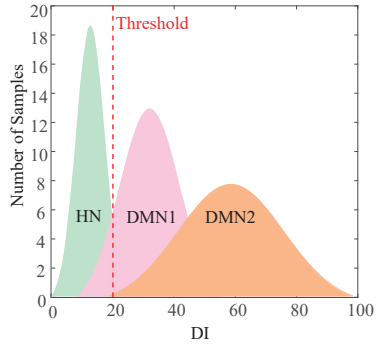

Fig. 28: DI distribution of PEH with $L = 0.34$m from DMN1 and DMN2 cases

*Appendix B: Impact of Sensing Location*

Since the sensors are often deployed as a network in the structure to achieve better detection efficiency, it is expected that the sensors have the ability to monitor and characterise the structural integrity across various locations. Note that the previous results are based on the same crack and sensing location, so this section studies the damage scenario (DMN1) as Section 5.1 but based on the sensing information collected at the quarter point. Fig.29 shows the damage characteristics of the acceleration signal under this scenario. Compared to the acceleration signal of DMN1 collected at mid-span, while the second mode dominates in the acceleration signal from quarter point for DMN1, visible changes from the first natural frequency under damaged conditions are also evident.

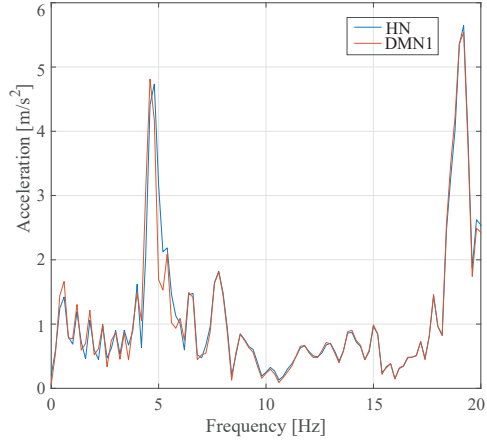

Fig. 29: Acceleration signals from the PEH at one quarter point under the DMN1 case

When 10% of the damage occurs at the mid-span, the results of energy harvesting and sensing utilising PEHs installed at the quarter position are illustrated in Fig.30. Specifically, Fig.30(a) reveals that the energy output in this scenario exhibits a similar trend to that observed in Fig.20(a), indicating a direct relationship between the energy harvesting performance ($E(L)$) and the installation location of the PEHs.

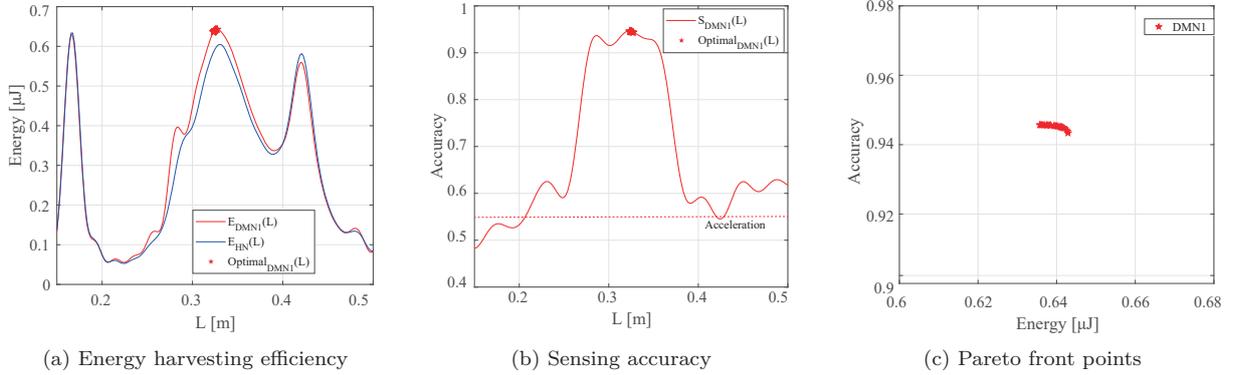

(a) Energy harvesting efficiency  (b) Sensing accuracy  (c) Pareto front points

Fig. 30: Bi-objective optimisation results of DMN1 with PEH installed at the quarter span

In addition, when compared with Fig.20(b) under the DQN1 scenario, the sensing accuracy depicted in Fig.30(b) exhibits contrasting changes across the PEH design space. For instance, the PEH with a fundamental frequency tuned to the second modal frequency of the bridge ($L = 0.17$m) cannot be effectively used for structural damage detection applications in this case. This limitation can be explained by the indistinguishable shift of the second fundamental frequency in Fig.31(a). As for the PEHs capable of harvesting energy in the first vibration mode (Fig.31(b, c)), the shift in the first fundamental frequency recorded in the voltage signal is valuable for damage detection. However, for the PEH with $L = 0.42$m (Fig.31(c)), the voltage signal collected at the quarter point is predominantly influenced by the second vibration mode, resulting in the second fundamental frequency as the primary component utilised in the training of the unsupervised learning models. Therefore, the shift of the first natural frequency under damage is mistaken for noise information, showing a lower sensing effect.

Rigorous NSGA-II optimisation reveals as a very narrow Pareto front, Fig.31(c), which can be seen a single point, corresponding to an optimal device with $L = 0.34$m and fundamental frequency of 4.8Hz, which matches the fundamental frequency of the bridge. Therefore, we can conclude that the installation position of the PEH has little effect on the optimal design for damage at mid-span.

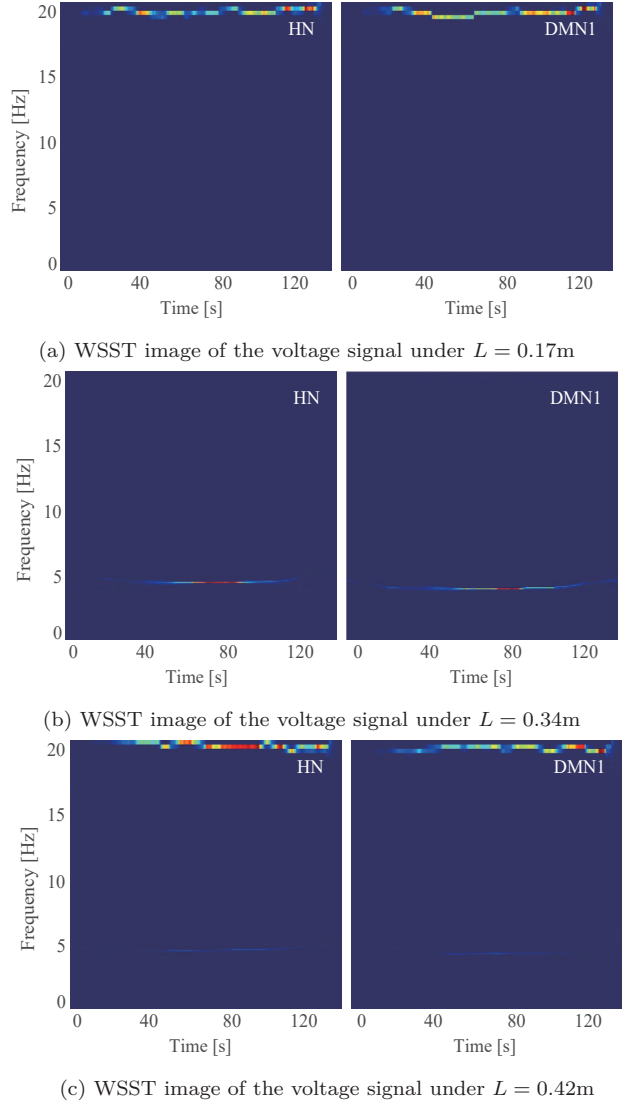

(a) WSST image of the voltage signal under $L = 0.17$m

(b) WSST image of the voltage signal under $L = 0.34$m

(c) WSST image of the voltage signal under $L = 0.42$m

Fig. 31: Illustration of the voltage signals of three PEHs with $L = 0.17$, $0.34$ and $0.42$m

*Appendix C: Impact of Damage Location and Road Roughness*

In this section, the application of PEH under 10% damage site at the quarter location (DQN1) is studied again, but based on the road B. Fig.32 shows the multi-objective optimisation results based on energy and energy per unit area, respectively, showing similar trends as in Section 5.2. However, compared to the results on road A (Fig.20), the optimal device under the rougher road surface are only concentrated around $L = 0.17$m, which are able to achieve excellent energy harvesting and signal filtering capabilities.

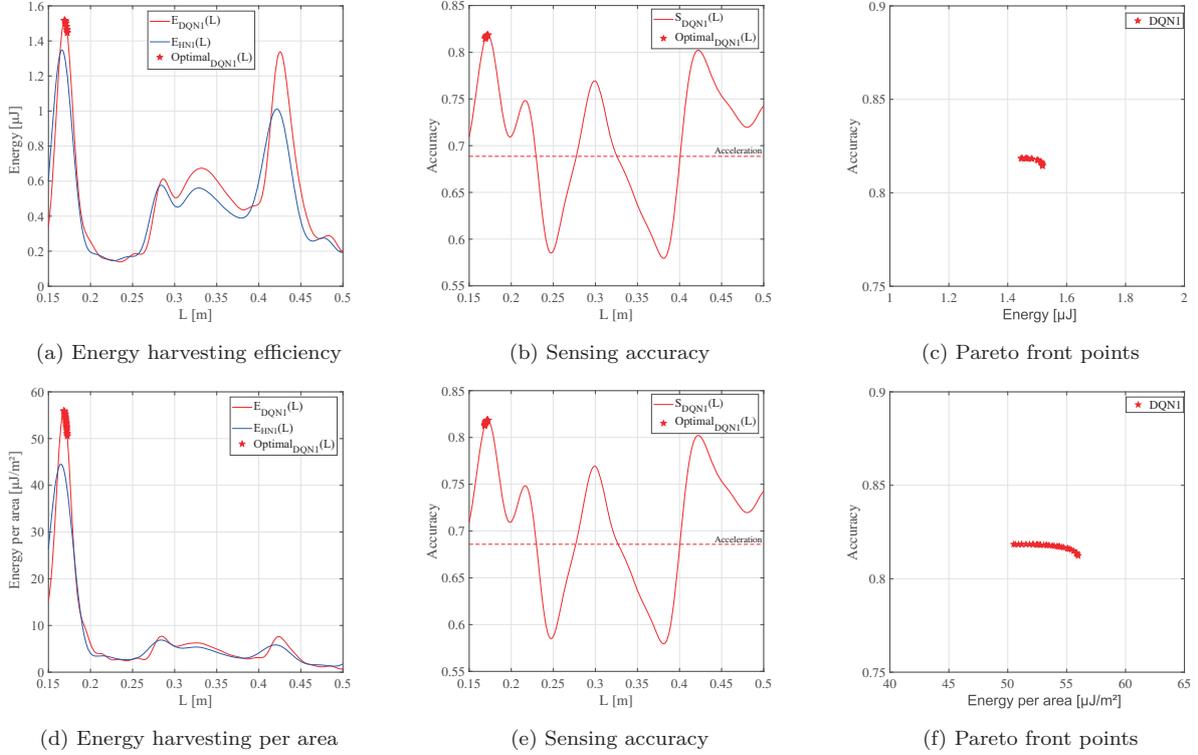

Fig. 32: Bi-objective optimisation results of DQN1 under Road B